\documentclass[%
 preprint,
 amsmath,amssymb,
 aps,
]{revtex4-2}

\usepackage{graphicx}% Include figure files
\usepackage{dcolumn}% Align table columns on decimal point
\usepackage{bm}% bold math
\usepackage{multirow}
\usepackage{comment}
\usepackage{url}        % For formatting URLs
\usepackage{hyperref}   % For clickable links
\usepackage[utf8]{inputenc}
\usepackage{setspace}
\begin{document}

\preprint{}

\title{From OpenSPIM to FlowSPIM: Enhancing the versatility of a light-sheet microscope through an iterative design process}% Force line breaks with \\

%\affiliation{%


\author{Endre Joachim Lerheim Mossige$^{1,2}$}
\thanks{endrejm@uio.no}
\author{Sergey Ponomartsev$^{3}$}
\author{Natalia Smirnova$^{3}$}
\author{Wietske van der Ent$^{4}$}
\author{Siri Andresen$^{6,7}$}
\author{Xian Hu$^{6}$}
\author{Felix Margadant$^{6}$}
\author{Rainer Heintzmann$^{9,10}$}
\author{Roland K\'ad\'ar$^{5,11}$}
\author{Kesavan Sekar$^{5,11}$}
\author{Helene Kn\ae{v}elsrud$^{6,7,8}$}
\author{Camila Vicencio Esguerra$^{4}$}
\author{Stefan Krauss$^{3}$}
\author{Dag Kristian Dysthe$^{2}$}
\author{Alexander Refsum Jensenius$^{1}$}

\affiliation{$^{1}$ RITMO Centre for Interdisciplinary Studies in Rhythm, Time and Motion, University of Oslo, Oslo, Norway}
\affiliation{$^{2}$ Njord Centre, Department of Physics, University of Oslo, Oslo, Norway}
\affiliation{$^{3}$ Hybrid Technology Hub, Faculty of Medicine, University of Oslo, Oslo, Norway}
\affiliation{$^{4}$ Chemical Neuroscience Group, Norwegian Centre for Molecular Biosciences and Medicine, University of Oslo, Oslo, Norway}
\affiliation{$^{5}$ Department of Mechanical Engineering, Chalmers University of Technology, 412 96 G\"oteborg, Sweden}
\affiliation{$^{6}$ Centre for Cancer Cell Reprogramming, Faculty of Medicine, University of Oslo, Oslo, Norway}
\affiliation{$^{7}$ Department of Molecular Medicine, Institute for Basic Medical Sciences, Faculty of Medicine, University of Oslo, Oslo, Norway}
\affiliation{$^{8}$ Department of Microbiology, Oslo University Hospital, Oslo, Norway}
\affiliation{$^{9}$ Leibniz Institute of Photonic Technology, Jena, Germany}
\affiliation{$^{10}$ Institute of Physical Chemistry and Abbe Center of Photonics, Friedrich-Schiller-Universit\"at Jena, Jena, Germany}
\affiliation{$^{11}$ Wallenberg Wood Science Centre (WWSC), Chalmers University of Technology, 412 96 G\"oteborg, Sweden}

%\date{}% It is always \today, today,

\begin{abstract}
 Light-sheet microscopy is ideal for the 3D imaging of large biological specimens, yet systematic documentation on adapting microscope designs to diverse experimental requirements remains limited. We present the iterative process used to design and build a modular light-sheet microscope and adapt it to different sample types and experimental conditions, including live and fixed zebrafish, Drosophila, and mouse gastruloids, as well as cellulose nanocrystals. Starting from a stripped-down OpenSPIM platform, we gradually made our microscope more versatile by adding hardware such as additional laser channels and optical filters, and new protocols for sample handling. In its final iteration, FlowSPIM is a new flow-based incubation technique for 3D tissue culture. %\textcolor{blue}{Text added since the initial submission in blue, although some new text is not colored blue.} \textcolor{red}{To-do or Referee's comments in red.}
\end{abstract}

\maketitle

%\tableofcontents

\section*{Introduction}
\label{sec:intro}
In light-sheet microscopy \cite{huisken2004optical, weber2014light, girkin2018light, stelzer2021light}, also known as Selective Plane Illumination Microscopy (SPIM), the sample is illuminated from the side by a micrometer-thin sheet of light aligned with the imaging system's focal plane. This means that out-of-focus regions of the sample between the light sheet and the focal plane do not contribute to the detected signal, yielding excellent optical sectioning performance, minimal photobleaching, and the ability to view millimeters into tissue. As such, SPIMs are ideal for 3D imaging of large samples such as zebrafish \cite{Keller2008ZebrafishDSLM,Huisken2009ReviewSPIM,Ahrens2013ZebrafishBrain} and fruit fly embryos \cite{Preibisch2008DrosophilaSPIM,Krzic2012Multiview,Schmied2016DrosophilaProtocol}, as well as laboratory-grown organoids including stem cell-based embryo models (SCBEMs) \cite{van2014symmetry,van20213d,ho2025spontaneous}. 

%Figure 1

SPIMs are also incredibly fast and can scan millimeter-sized samples in seconds to characterize cell migration in 3D \cite{huisken2004optical}. Because only a thin section of the sample is illuminated, SPIMs have extremely low phototoxicity compared with more conventional techniques such as confocal microscopy, helping maintain the viability of fragile, living samples \cite{weber2014light}. Beyond these advantages, their modular design enables swapping optical components, pumps, and temperature-control devices for targeted applications. Starting from the OpenSPIM platform \cite{pitrone2013openspim,webpagekeyOpenSPIM},~ \citet{girstmair2022time} added a second detection objective to image flour beetles from two sides simultaneously. By interfacing a home-built SPIM with an environmental chamber, ~ \citet{lorenzo2011live} demonstrated long-term imaging of tissue culture. Building on this setup, ~ \citet{pampaloni2014tissue} showed that Madin-Darby canine kidney (MDCK) tissue cultures can remain viable longer by continually exchanging the cell media. 

\begin{figure}[t]
    \centering
    \includegraphics[width=0.8\linewidth]{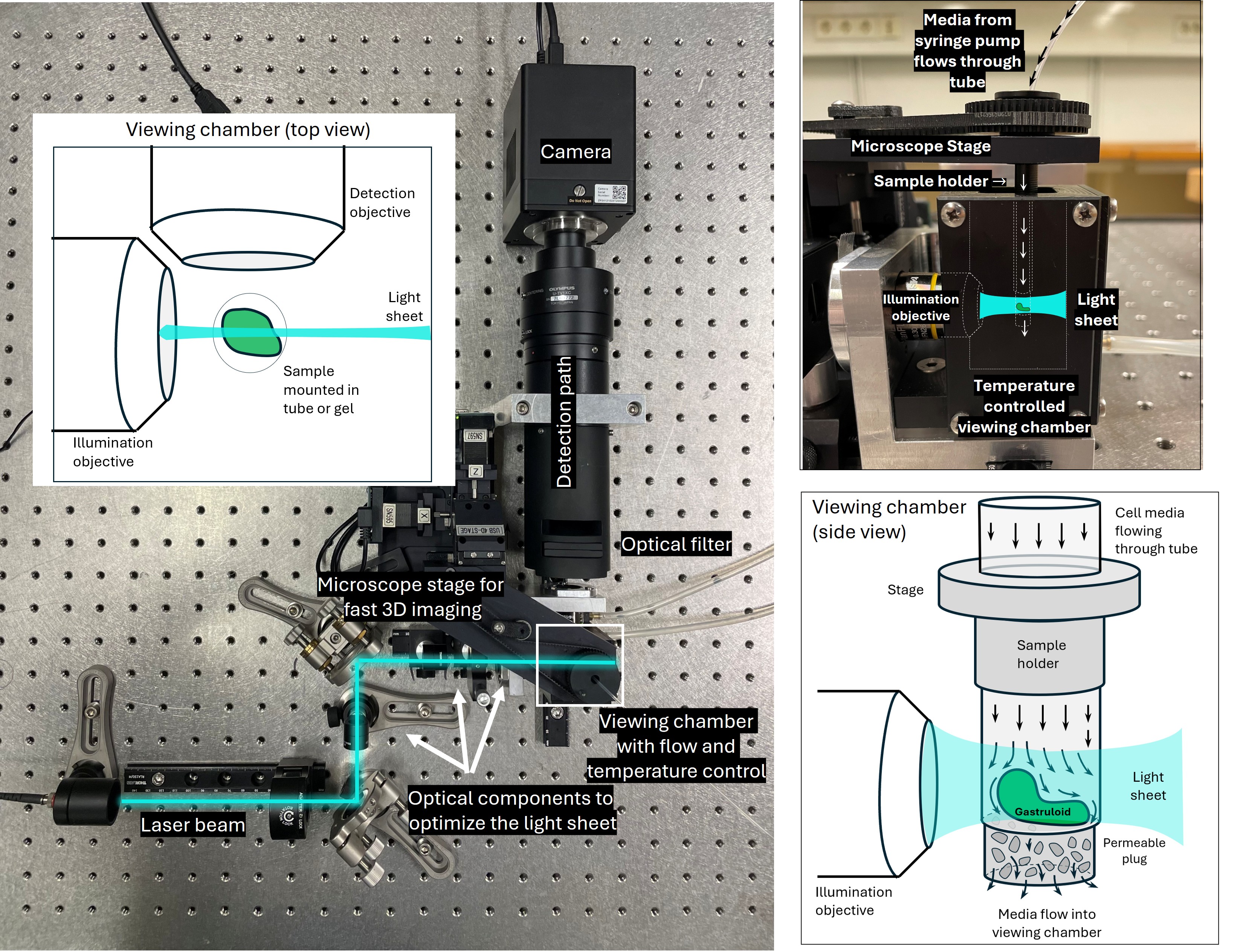}
    \caption{Key features of the latest generation of the setup, FlowSPIM. An incubator provides physiological temperature and flow conditions to a 3D tissue culture, such as a mouse gastruloid, inside a capillary tube. A micrometer-thin light sheet illuminates the sample from the side, and the optical cross-sections are reconstructed into sharp 3D images.}
    \label{fig:lastversionofsetup}
\end{figure}

However, while existing reports have demonstrated applications across multiple imaging modalities and diverse biological contexts, more comprehensive documentation is needed on systematically adapting microscope designs to different sample types and experimental requirements. In addition, the iterative design process, describing how to build a light sheet microscope step-by-step and adjusting it to specific needs, is not well documented. Here, we document the iterative design process for building a stripped-down light-sheet microscope and how we built on the initial design by adding new hardware, such as additional laser lines, different optical filters, an incubation chamber, and new sample-handling protocols. Our goal has been to characterize how live stem cell-based embryo models, so-called gastruloids\cite{BaillieJohnson2015}, grow and develop in 3D, as shown in Fig. \ref{fig:lastversionofsetup}. Towards this goal, we image zebrafish and fruit fly (Drosophila) larvae in 3D, and we demonstrate label-free light-sheet 3D imaging of cellulose nanocrystals for the first time. For each setup generation, we demonstrate new functionalities, discuss limitations, and present best-practice experimental procedures. We start by describing the iterative design process. 

\section*{The iterative design process}
\begin{figure}[p]
    \centering
    \includegraphics[width=0.8\linewidth]{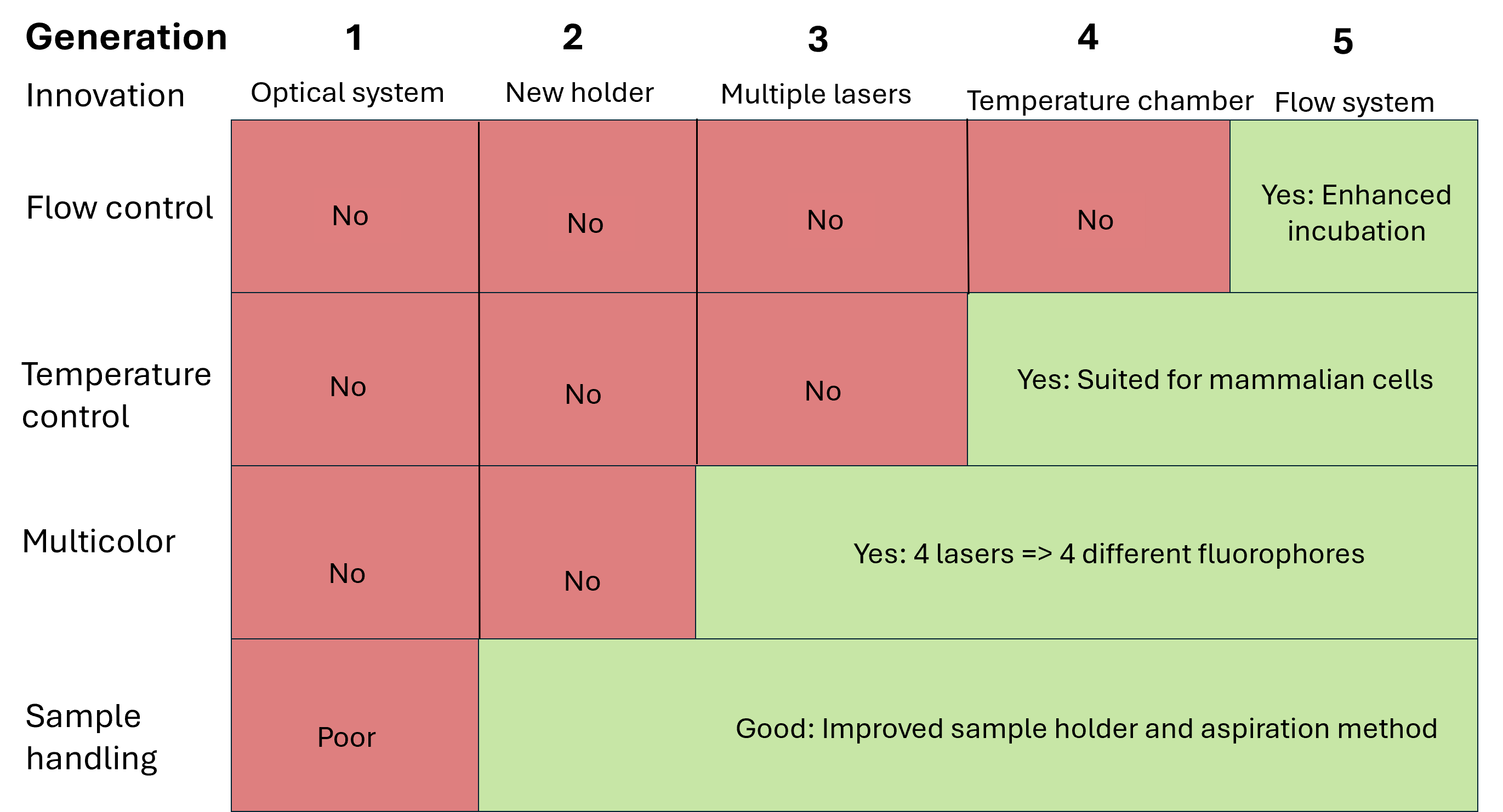}
    \caption{The schematic shows the iterative design process. The main innovation of Generation 1 is the optical system of the light-sheet microscope. The improved sample holder and mounting technique in Generation 2 significantly enhances handling. The new laser in Generation 3 enables multicolor imaging of up to four different fluorescent markers. The temperature chamber in Generation 4 enables incubation of mammalian tissue culture. Finally, Generation 5, the FlowSPIM, features flow control to ensure efficient transport of oxygen and nutrients to mammalian cell cultures, enhancing viability.}
    \label{fig:iterativedesign}
\end{figure}

Figure \ref{fig:iterativedesign} and Table \ref{tab:iterativedesign} provide an overview of the five setup generations and their functionalities. Starting with the OpenSPIM platform, Generation 1 is a simple SPIM with a single (blue) illumination laser and no temperature or flow control. Section \ref{sec:generation1} describes the iterative process of building four subsequent prototypes of the optical system, where each prototype addresses the limitations of the previous one to maximize optical performance, user-friendliness, and ease of laser alignment. In Generation 2, we improve the sample holder and mounting technique, and Section \ref{sec:generation2} describes how this enables 3D imaging of zebrafish, Drosophila, gastruloids, and liquid crystals. In Generation 3, described in Section \ref{sec:generation3}, we add three laser lines (UV, green, and red) to the initial blue laser to visualize cells marked with fluorescent reporter proteins, such as the various germ layers in gastruloids. In Generation 4, we add temperature control to enable live imaging of mammalian tissue culture, and Section \ref{sec:generation4} describes how we designed, built, and evaluated three iterations of our temperature control unit. Finally, in Generation 5, presented in Section \ref{sec:generation5}, we propose a fully automated, flow-based incubation technique, the FlowSPIM, which facilitates efficient transport of nutrients and oxygen to 3D gastruloids to enhance their long-term viability and survival. We present useful troubleshooting strategies, discuss how to improve our setup, and place our work in a scientific context in Section \ref{sec:discussion}. We summarize our findings and conclude in Section \ref{sec:conclusion}.  

%Figure 2

%Table 1

\section{Generation 1: Optical setup}
\label{sec:generation1}
\begin{figure}[]
    \centering
    \includegraphics[width=1.0\linewidth]{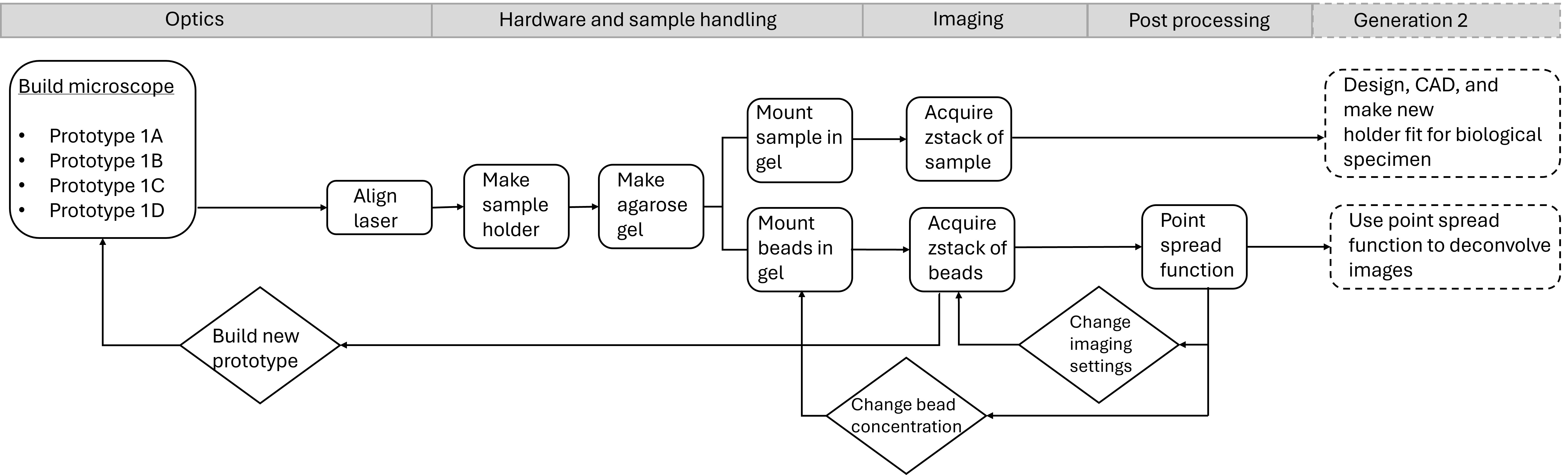}
    \caption{Iterative design process for building the optical system in Generation 1 of the setup. The setup went through four iterations (Prototypes 1A--1D) to optimize optical performance and ease of laser alignment. We used the point spread function, obtained by imaging a stack of fluorescent beads suspended in an agarose gel, to quantify the optical system's resolution and improve images of biological specimens through deconvolution in subsequent generations of the setup.}
    \label{fig:flowchartGen1}
\end{figure}

Figure \ref{fig:generation1setup_11mai2026} shows the four prototypes (1A-1D) of Generation 1 of our setup, developed through an iterative process, as illustrated by the flowchart in Figure \ref{fig:flowchartGen1}. In subsequent Sections, we describe how we built and evaluated each prototype, starting with prototype 1A.  

\subsection{Prototype 1A: OpenSPIM} 
\begin{figure*}[t]
    \centering
    \includegraphics[width=1.0\linewidth]{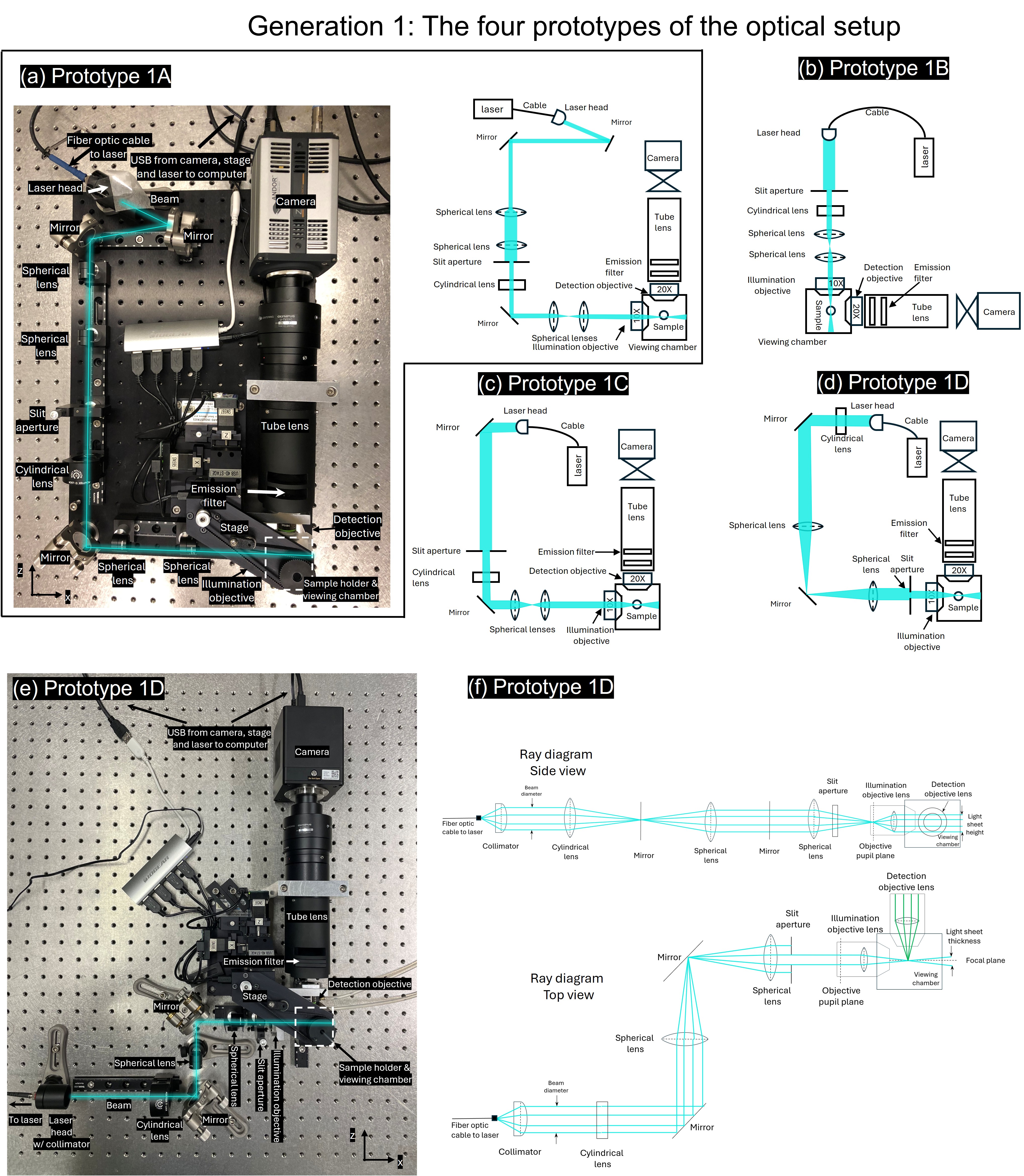}
    \caption{Figs. (a)-(d) show the four different prototypes of the optical setup in Generation 1. \textbf{(a)} Prototype 1A is built based on the OpenSPIM platform \cite{pitrone2013openspim}. \textbf{(b)} Prototype 1B is identical to Prototype 1A, except that the mirrors were removed and the first telescope was replaced with a laser head with a built-in collimator. \textbf{(c)} Prototype 1C is identical to Prototype 1B, but adds two alignment mirrors. \textbf{(d,e)} In Prototype 1D, the cylindrical lens is placed before the vertical slit aperture. This slight modification drastically improves the light sheet's properties and makes laser alignment much easier. To reduce the phototoxic dose for living species and photobleaching, one could add a neutral-density filter and a field aperture to the illumination path. \textbf{(f)} Ray diagrams showing the illumination path for Prototype 1D.}
    \label{fig:generation1setup_11mai2026}
\end{figure*}

\begin{figure*}[]
    \centering
    \includegraphics[width=1.0\linewidth]{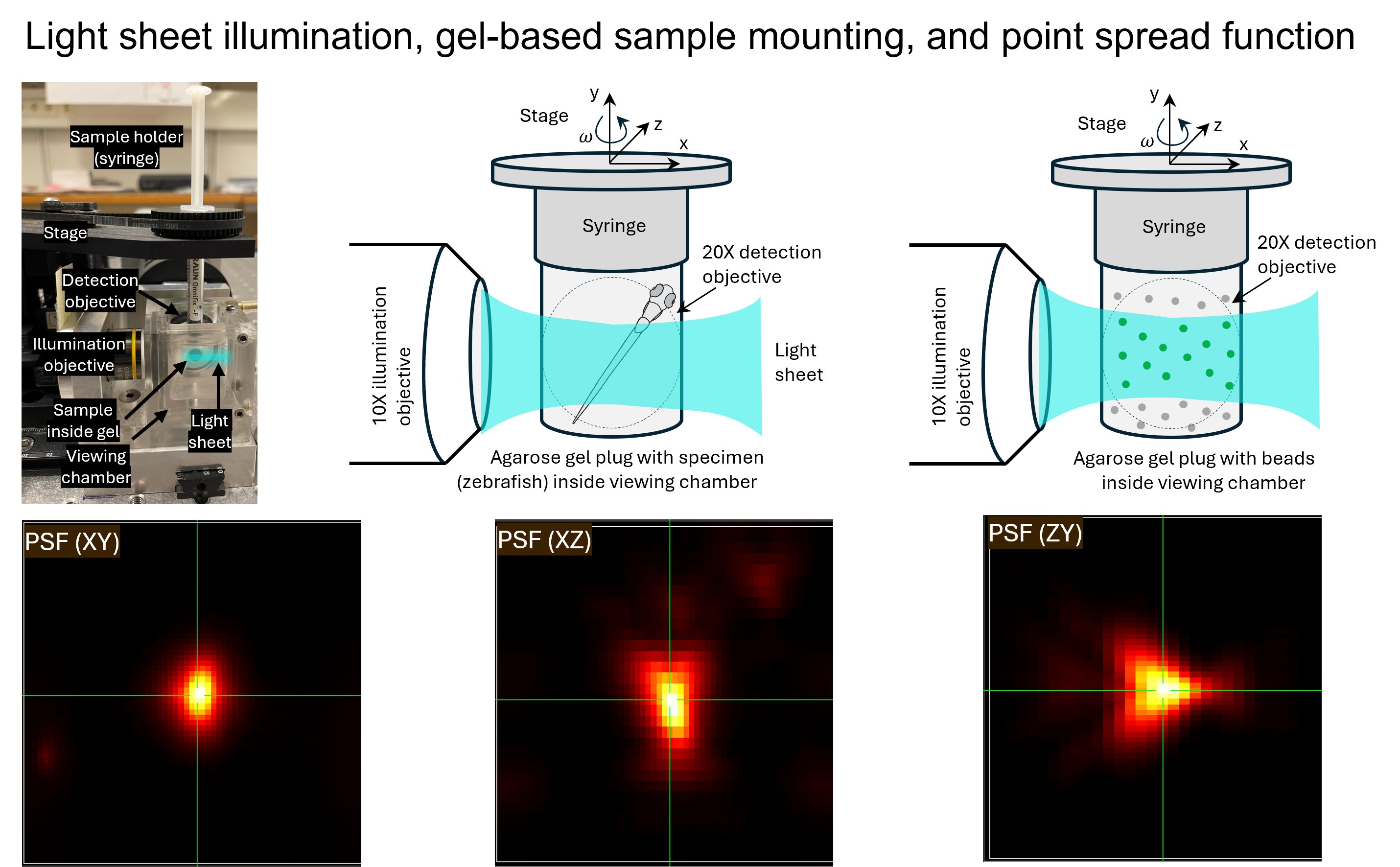}
    \caption{\textbf{Top row:} Side views showing illumination optics inside the water-filled viewing chamber in Generation 1 of the setup. The light sheet comes from the left through the 10$\times$ water-dipping illumination objective, illuminating, for example, a biological specimen such as a zebrafish (middle) or fluorescently tagged beads (right). In Generation 1, the sample of interest is embedded in a freely hanging agarose cylinder, whereas in subsequent generations, it is mounted inside a plastic tube. A 20$\times$ water-dipping detection objective is placed perpendicular to the illumination objective, as indicated by the dotted circle in the middle and right figures. \textbf{Bottom row:} Point spread function (PSF) in the final optical configuration (Prototype 1D), obtained from images of fluorescent beads embedded in agarose and illuminated with blue laser light ($\lambda$=488 nm). Left: xy-plane. Middle: xz-plane. Right: yz-plane.}
    \label{fig:Gen1setup_illuminationGelbasedMounting}
\end{figure*}

\textbf{Illumination optics:} Prototype 1A was built based on the OpenSPIM platform \cite{pitrone2013openspim}, following the protocols in Ref. \cite{webpagekeyOpenSPIM}, see Figure \ref{fig:generation1setup_11mai2026}(a). An optically pumped laser (OBIS 488 LX, Coherent; beam diameter: 0.8 mm and maximal power: 50 mW, where the power is adjusted with Coherent Connection from Coherent) delivers blue light (wavelength $\lambda$=488 nm), and a fiber-optic cable (PMC-400RGB-3-18E/28E-200, Schäfter+Kirchhoff) guides it to a laser head clamped to the optical table (using a 20 mm post (TR20/M, ThorLabs), a 25 mm post holder (PH20E/M, ThorLabs), a Ø1$\text{''}$ lens mount (SMR1/M, ThorLabs), a fiber adapter plate (SM1FCA, ThorLabs), and a clamping fork (CF175, ThorLabs)). Two alignment mirrors (BB1-E02, Ø1$\text{''}$, ThorLabs; mounted onto Polaris KA Side Retention Mounts for Ø1$\text{''}$ Optics, ThorLabs) guide the beam through two spherical lenses (AC127-050-A-ML, f = 50 mm; AC127-025-A-ML, f = 25 mm, both ThorLabs; spaced 75 mm apart) to collimate the light and expand the beam area, acting as a beam-expanding telescope. After the telescope, a vertically oriented, adjustable slit aperture (VA100/M, ThorLabs) shapes the beam into a preliminary light sheet, and a cylindrical lens (ACY254-050-A, f = 50 mm, ThorLabs) focuses the light sheet onto a third alignment mirror (BB1-E02, Ø1$\text{''}$, ThorLabs; mounted onto Polaris KA Side Retention Mounts for Ø1$\text{''}$ Optics, ThorLabs). The beam then passes through two spherical lenses (AC127-050-A-ML, f = 50 mm, ThorLabs; AC127-025-A-ML, f = 25 mm, ThorLabs; spaced 75mm apart) whose purpose is to relay the light sheet onto the objective pupil plane of a water immersed illumination objective (UMPLFLN 10XW, 10X/0.5, Olympus), mounted to a custom made (milled) water filled acrylic chamber (inner dimensions: width of 22 mm, length of 22 mm, height of 37 mm). A spherical collimation lens inside the immersion objective rotates the light sheet by 90 degrees from horizontal to vertical and focuses it onto the sample (e.g., a biological specimen or fluorescent beads); see Figure \ref{fig:Gen1setup_illuminationGelbasedMounting}. Finally, a neutral density (ND) filter (ND507A, ThorLabs; optical density: 0.7, transmission $\sim$ 20$\%$; not shown) was placed between the two first spherical lenses (can be placed anywhere in the light path) to reduce the phototoxic dose on the biological specimen. We used the ND filter for the majority of the experiments presented in this paper. 

Following the OpenSPIM protocol, the sample is embedded in an agarose cylinder, freely hanging from a standard 1\,mL syringe (Omnifix F, B-Braun), cut to length using a wallpaper knife; see Figure \ref{fig:holders}(a). A movable stage (USB-4D-Stage, Picard Industries) holds the syringe and interfaces with a lab computer via Micro-Manager 2.0.0\cite{edelstein2010computer}, allowing the operator to rotate and translate the sample along the x, y (up/down), and z axes. 

\textbf{Detection system:} The detection objective (UMPLFLN 20$\times$W, 20$\times$/0.5, Olympus) sits perpendicular to the illumination objective in the viewing chamber; see Figure \ref{fig:generation1setup_11mai2026}(a) and Figure \ref{fig:Gen1setup_illuminationGelbasedMounting}.
A multi-bandpass emission filter (Chroma ZET 488/561m) sits in one of the two slots on an infinity space tube (self-made; for CAD drawings see Ref.\cite{webpagekeyOpenSPIM_parts}) between the detection objective, tube lens (U-TLU, Olympus; f=180 mm, M=1$\times$) and light-sensitive sCMOS camera (Andor Zyla 5.5; 5.5 megapixels, cooled sCMOS sensor) and removes reflections and scattering by blocking wavelengths outside the transmission bands (503nm - 546nm and 576nm-710nm). The imaging system voxel size (x, y, z) is 0.325 $\mu$m × 0.325 $\mu$m × 1.542 $\mu$m. Here, the image plane dimensions (x, y) are set by the camera pixel size (6.5 $\mu$m/pix) and the detection objective magnification (M=20$\times$), and the z-dimension is set by the 4D microscope stage described above. The tube lens does not affect voxel size because it has no magnifying or demagnifying optics (M=1$\times$).  

\textbf{Characterization:} To assess the quality of the light sheet, we imaged fluorescent beads (F-Y 050, Estapor Microsphères; mean diameter: 0.5 $\mu$m; Acquisition settings: Laser power of 10 mW, exposure time of 1000 ms) embedded in a low gelling point agarose gel (Sigma-Aldrich; 1\% concentration), see Figure \ref{fig:Gen1setup_illuminationGelbasedMounting}. We also used these imaging parameters in subsequent Generation 1 prototypes (B-D) to assess light-sheet quality.

\textbf{Evaluation:} We could not generate a thin, flat light sheet, as shown by the grayscale intensity of the beads across the field of view in the microscope images. In addition, we found that using as many as three individual mirrors to align the laser was overly complicated, and we were unable to produce truly collimated light with the beam-expanding telescope. 

\subsection{Prototype 1B: Removing the alignment mirrors and the first telescope}
In Prototype 1B, we simplified the setup by removing the three alignment mirrors (BB1-E02, Ø1$\text{''}$, ThorLabs) and by replacing the first telescope system (made with two spherical lenses) with a Ø1$\text{''}$ aspherical lens (AL2520, ThorLabs; f=20 mm) mounted inside of a lens tube (SM1L10, Thorlabs) as a means to collimate the light, see Figure \ref{fig:generation1setup_11mai2026}(b). 

\textbf{Evaluation:} Using the same imaging parameters as for Prototype 1A (Laser power: 10 mW, exposure: 1000 ms) to assess the quality of the light sheet, we found that replacing the beam-expanding telescope with the collimation optics on the laser head was an efficient way of collimating the light. However, we could not align the laser solely by turning the laser head side to side and found that micro-adjustments from the alignment mirrors were necessary. 
%In addition, the laser head can not be rotated in the vertical direction.

\subsection{Prototype 1C: Bringing back the alignment mirrors}
In Prototype 1C, see Figure \ref{fig:generation1setup_11mai2026}(c), we brought back two of the alignment mirrors from Prototype 1A, which significantly simplified laser alignment. However, because of the light sheet's high (Gaussian) curvature, we could not obtain a flat light sheet across the detection system's field of view. We used the same imaging parameters as for Prototype 1A/B to assess the quality of the light sheet.

\subsection{Prototype 1D: Improving the light sheet uniformity by placing the aperture after the cylindrical lens}
In Prototype 1D, we placed the slit aperture after the cylindrical lens instead of before it to obtain a more uniform light sheet across the field of view; see Figure \ref{fig:generation1setup_11mai2026}(d,e,f). In this configuration, the cylindrical lens focuses the beam into a horizontal line (sheet) on the first mirror, and a 4$\textit{f}$ relay system made from two spherical lenses (ACY254-050-A, f = 50 mm, ThorLabs; placed 2f=100 mm apart) relays (copies) this line onto the objective pupil plane of the illumination objective, where a collimation lens flips the light sheet from horizontal to vertical and focuses it onto the specimen. The adjustable slit aperture sits between the last spherical lens and the illumination objective and crops the horizontal line made by the cylindrical lens, efficiently reducing the curvature of the light sheet in the propagation (x) direction. By gradually closing the aperture (thereby lowering the numerical aperture for the light entering the lens), the light sheet becomes increasingly flatter (with the small expense of increasing its waist (minimal thickness)), ideally yielding a uniform thickness across the field-of-view of the camera (1 mm by 1 mm). By imaging 0.5 $\mu$m fluorescent beads embedded in a freely hanging agarose cylinder, we found a slit width of 2 mm to be a good compromise between light sheet uniformity and thickness.

\textbf{Resolution of the optical system:} The x-y resolution of the optical system is 1.5 $\mu$m by 2.1 $\mu$m (in the image plane) by 6.0 $\mu$m (along the optical (z) axis), as measured by the full width at half maximum (FWHM) of the point spread function (PSF). Here, the z-direction resolution can be used as a measure of the light-sheet thickness. The PSF was obtained by first acquiring an image stack of 0.5~$\mu$m fluorescent beads (laser power: 10 mW, exposure time: 1000 ms, z step length: 1.542 $\mu$m, number of images: 65) and then using the PSF distiller in Generic Deconvolution (in-house code developed by one of the authors (RH), available upon request). In subsequent generations of the setup, we subtracted the PSF from the raw images through deconvolution using either Generic Deconvolution or Huygens Scientific Volume Imaging to reduce blur and to mitigate photon noise. 

\textbf{Sample holder and embedding technique:}
While a simple sample holder and gel-based mounting technique suit fluorescent beads, they are incompatible with biological samples, especially gastruloids. The reason for this is threefold. First, the syringe does not fit snugly in the stage, so it slips during sample rotation, such as when the sample is imaged from different sides. Second, since the agarose cylinder is much wider than the biological sample, it is difficult to align it with the stage's rotation (y) axis; see Figure \ref{fig:Gen1setup_illuminationGelbasedMounting}. This causes the sample to translate during rotation, sometimes out of the camera's field of view. Third, our gastruloids need to be in liquid cell culture media rather than a stiff gel to grow unhindered, aligning with established protocols; see e.g.~\citet{BaillieJohnson2015}.  

\section{Generation 2: Improving the sample holder and mounting technique} 
\label{sec:generation2}

\textbf{Overall design criteria:} 
To overcome the limitations of the sample holder and embedding technique described above, we propose enclosing the sample and cell media in a capillary tube, sealed from below to prevent leakage into the viewing chamber. To minimize light scattering and optical distortions from spherical aberration, the capillary should have a refractive index that closely matches water. The capillary walls must also be thick enough to prevent bending during mounting and sample handling. In addition, to prevent the sample from translating out of the field of view during rotation when imaging from different sides, the capillary should be straight, and its inner diameter should fit within the camera's field of view (1~mm by 1~mm). A small inner diameter also ensures that long, slender specimens, such as Drosophila and zebrafish larvae, stand upright in the tube, aligning with the microscope stage's rotation (y) axis. Finally, to facilitate slip-free rotation during imaging from different sides, the capillary must fit snugly in the capillary holder, which in turn must fit snugly in the microscope stage.

\subsection{Prototype 2A: Modifying the syringe-based tube holder}
Following the guidelines provided in Refs. \cite{samplemounting,kaufmann2012multilayer}, we mounted our samples using syringe and needle aspiration (Omnifix-F Luer Solo, B-Braun) in fluorinated ethylene propylene (FEP) tubes, since these tubes minimize optical distortions because their refractive index (1.34) closely matches that of water (1.33). A wide range of inner and outer diameters is available for such tubes, and we chose a tube (BOLA) with an inner diameter of 0.8~mm to minimize rotation-induced translation of the sample. The thick walls (0.4~mm) ensure mechanical stability during sample handling and mounting in the tube holder. 

Figure \ref{fig:Gen2overview} (a) and Figure \ref{fig:holders} (b) show the first prototype of the tube holder and mounting method. As in Generation 1, the tube holder is made by cutting an empty plastic syringe (Omnifix-F Luer Solo, B-Braun) with a wallpaper knife. Instead of mounting the sample in agarose and aspirating it with the newly cut syringe, we inserted a Teflon plug with a central hole (diameter: 0.8\,mm) into the syringe to hold the FEP tube in place. To seal the tube, we gently pushed it into an agarose slab prepared by leaving a 1\% agarose solution to dry in a Petri dish. Once the sample is mounted, we cut the tube with a razor blade and place it in a tube holder attached to the microscope stage in front of the imaging objective.  

To evaluate the sample holder, tube, and sealing method, we mounted fixed zebrafish larvae (length $\sim$5 mm, width $\sim$0.7 mm) into the FEP tube using syringe and needle aspiration and performed z-stack imaging (laser power: 20 mW, exposure time: 200 ms, z step size: 1.542 $\mu$m). Immediately, we noticed severe striping artifacts. These artifacts did not originate from the sample, as often reported in the literature \cite{Huisken2007mSPIM}, but instead came from the tubing. We also found it difficult to align the tube with the stage's rotation axis because the tubes were not straight and because the Teflon plugs were skewed in the holder. Finally, we could not create flat agarose plugs to seal the tube because the agarose cracked when we pushed the tube into the solidified agarose solution.    

%Figure 3

\subsection{Prototype 2B: Improved sample holder and mounting method}
\textbf{Cleaning, straightening, and sealing the tubes:}
Inspired by established protocols \cite{samplemounting}, we cleaned the tubes before each imaging experiment by flushing them with lye solution, ethanol, and de-ionized (DI) water, before ultrasonicating them for 15 minutes. To straighten the tubes, we pushed them into circular glass capillaries, placed them in water-filled laboratory flasks (200\,mL), and microwaved them for 3 minutes. Finally, after aspirating the sample and media, we gently pushed the tubes into a synthetic Sigillum wax sealant (LW-Critwax, LW Scientific).    

\textbf{Improved sample holder:}
We fabricated an improved micromachined sample holder by lathe turning; see Figure \ref{fig:Gen2overview}(b,c,d). Importantly, the O-ring sitting on the top part of the holder ensured a tight fit with the stage.   

\textbf{Evaluation of the sample holder and mounting technique:}
To evaluate Prototype 2B of our setup, we mounted fixed zebrafish, fixed gastruloids, and live Drosophila larvae in the tube using syringe- and needle-based aspiration. We sealed the tube with the wax sealant as described above, and performed z-stack imaging from different sides. 

\textit{Fixed zebrafish:} Figure \ref{fig:Gen2overview}(e) is an image of a fixed zebrafish larva (6 days post fertilization; 6 dpf) mounted in the FEP-tube, and Figure \ref{fig:Gen2overview}(f) is an optical section of the blood vessel network inside the brain of such a sample (dorsal view; laser power: 20 mW, exposure time: 200 ms, z step size: 1.542 $\mu$m). Because of the tube confinement, the sample's long axis aligns with the rotation axis, effectively minimizing sample translation during rotation on the microscope stage. The tube-straightening step and the new tube holder further minimized translation during sample rotation. Finally, cleaning the tubes efficiently removed the striping artifacts, and the synthetic wax plug provided a tight seal.

\textit{Fixed gastruloids:} While mounting single, fixed zebrafish in the tube with standard syringe- and needle-based aspiration worked well, it did not work as well for gastruloids, as we could not aspirate individual samples. Instead, we typically aspirated 3-4 samples into the tube simultaneously. Also, the wax plug blocked the light sheet so that the samples were only partially illuminated. 

\textit{Live Drosophila:} Our ultimate goal was to perform 3D imaging of live gastruloids. As such, a natural stepping stone was to perform live imaging of model organisms, as they do not require an incubation chamber (to be introduced in Generation 4 and 5 of our setup). To this end, we acquired live Drosophila larvae. To facilitate mounting in the tube, we arrested their motion by exposing the larvae to ether gas (an anesthetic) before aspiration. This allowed us to image live samples for 10 minutes (laser power: 10 mW, 500 ms exposure), but after that the anesthetic was no longer effective, causing the larva to move and making imaging impractical.   
 
\textbf{Conclusion:} The improved sample holder, combined with the tube straightening and cleaning step, facilitated imaging of fixed zebrafish. However, the sample mounting technique and sealing method were impractical for the imaging of fixed gastruloids and live Drosophila larvae. In Prototype 2C described below, we aim to improve the mounting technique for live model organisms such as Drosophila and zebrafish, and in Prototype 2D we aim to improve the mounting method for gastruloids.

\subsection{Prototype 2C: Gel-based mounting for live model organisms}

To arrest the motion of live zebrafish and Drosophila larvae during imaging, we mounted them in a stiff, low-gelling-temperature agarose gel inside the capillary after anesthetizing them. 

\textbf{Protocol:} First, we anesthetized Drosophila with ether gas, and zebrafish with tricaine solution to arrest their motion. Second, we melted a 1\% agarose solution on a hot plate, lowered the temperature to 40$^\circ$\text{C} (measured with a thermocouple (Standard ST-612, Clas Ohlson)), and transferred the samples by pipetting. Third, we used a syringe and needle to aspirate the sample and agarose solution into the FEP tube. Finally, we waited five minutes for the agarose to solidify, then mounted the sample in the tube holder and attached it to the microscope stage in front of the camera.

\textbf{Evaluation:} While the 1\% agarose gel restricted the motion of model organisms, it did not completely arrest it. As such, we made the agarose gel stiffer by increasing the concentration to 4\% and found that it completely arrested motion, making it possible to acquire image stacks of individual neurons in zebrafish and fat body tissue in Drosophila larvae from different angles. We ensured the zebrafish remained viable throughout the imaging experiment by monitoring their heart rate with brightfield imaging, which we enabled by turning on the room lights. Finally, to obtain the 3D image in Figure \ref{fig:Gen2overview}(g), we combined the z-stacks of Drosophila (laser power: 10 mW, exposure time: 500 ms) imaged from 18 different angles using Huygens' multi-view fusion algorithm, and further improved image quality by reducing the influence of photon noise through deconvolution using the point-spread function measured in Generation 1. 

\subsection{Prototype 2D: Improving the mounting technique for small samples and cellulose nanocrystal suspensions}

To aspirate single gastruloids into the tube, the syringe and needle must be completely free of air. As such, conventional luer slip-tip syringes and needles (top panels in Figure \ref{fig:mounting}) cannot be used due to their inherent dead volumes. Instead, we used luer-lock syringes interfaced with flanged tubing via male-female connections, which have minimal dead volume, to aspirate single gastruloids into the tube (bottom panels in Figure \ref{fig:mounting}). To avoid the seal blocking the light sheet, we used a steel pin (diameter = 0.78 mm) to seal the tube from below, replacing the soft synthetic polymer used in previous iterations. 

\textbf{Building an air-tight aspiration system:} To make an air-tight aspiration system, we first threaded the male and female connectors (Flanged Fittings Kit, IDEX) onto the tube, and used a flange kit (Easy-Flange, VICI) to make flanged tubing. We then brought the flange into contact with the male-female connector, and screw the connectors onto a luer lock syringe (Omnifix, B-Braun), see the bottom panels in Figure \ref{fig:mounting}. Next, we aspirated a single gastruloid, let it sediment, and inserted the steel pin to seal the tube. To prevent the sample from sticking to the tube wall, we pre-rinsed the tube with a surfactant solution (Anti-Adherence Rinsing Solution, Stemcell Technologies).  

\begin{figure*}[]
    \centering
    \includegraphics[width=1.0\linewidth]{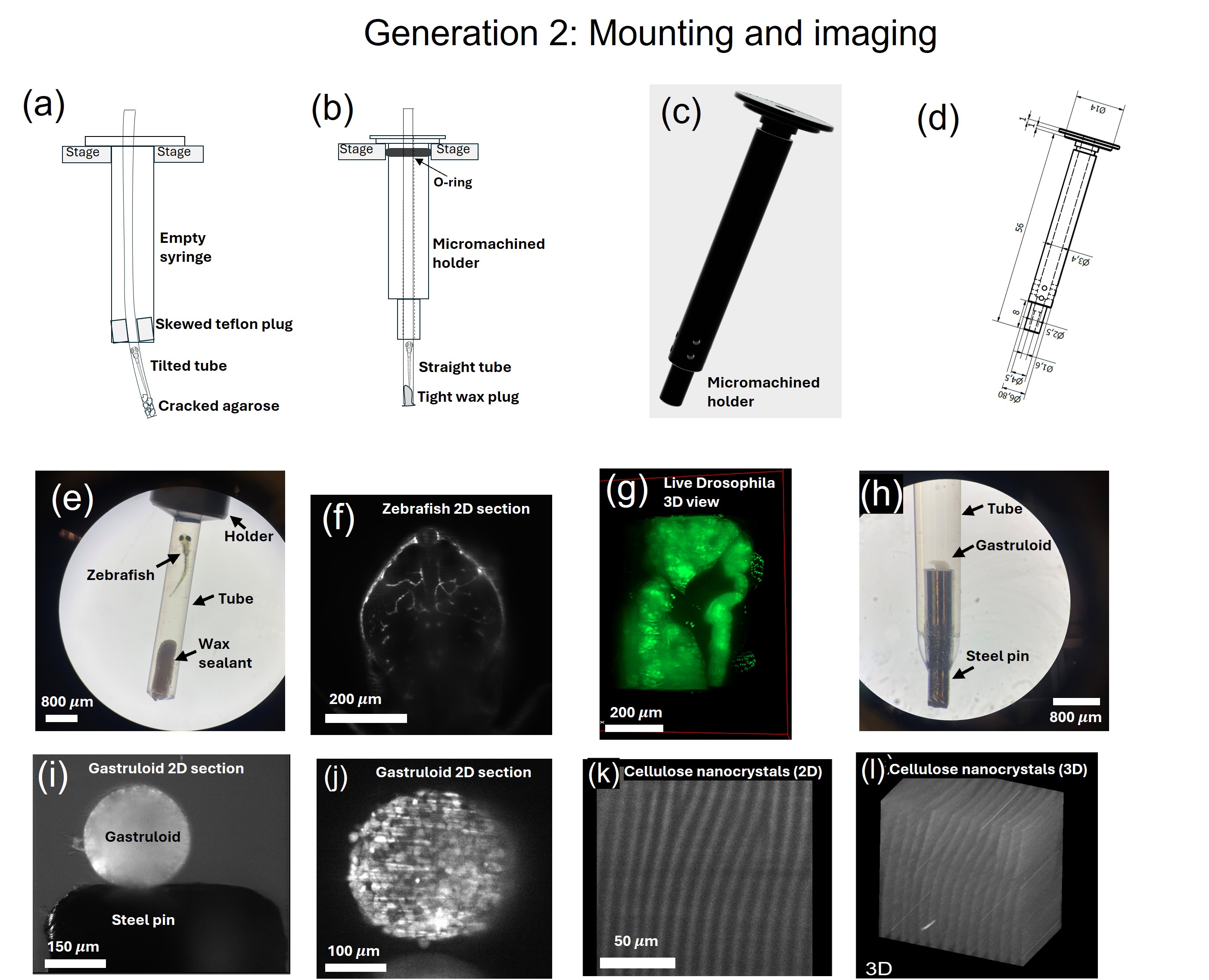}
    \caption{Mounting and imaging of zebrafish, Drosophila, gastruloids, and a suspension of biphasic chiral nematic liquid crystals (cellulose nanocrystals (CNCs) dispersed in water) in Generation 2 of the setup. \textbf{(a)} Prototype 2A: Sample holder made from an empty syringe. A Teflon plug with a hole in the middle holds the tube in place, but because the plug does not fit snugly in the syringe, the tube tilts relative to the rotation axis. \textbf{(b)} Prototype 2B: With the improved micromachined (lathe-turned) sample holder, the sample aligns with the straightened tube, thereby minimizing translation during sample rotation. \textbf{(c)} 3D CAD drawing and \textbf{(d)} 2D mechanical drawing of the micromachined holder. \textbf{(e)} Model organisms such as zebrafish are mounted in the tube with syringe and needle aspiration, and the tube is sealed with a soft wax sealant. Live samples are mounted in agarose gel without the polymer plug (not shown). \textbf{(f)} Maximum intensity projection made from ten optical sections of GFP-tagged blood vessels in the brain of a zebrafish larva (6 days post fertilization, dpf). \textbf{(g)} 3D rendered and deconvolved fat body tissue expressing GFP-Atg8a in a Drosophila larva. \textbf{(h)} Gastruloids are aspirated into the tube using an airtight Luer-lock syringe and flanged fitting, and the sample rests on a stainless steel pin. \textbf{(i,j)} Optical slices showing T/Bra GFP (mesoderm) reporters in a mouse gastruloid (3 days post-aggregation, dpa). \textbf{(k)} Optical section of bands of untagged mesoscale CNC self-assembled structures. \textbf{(l)} 3D rendered view of the same CNC structures.}
    \label{fig:Gen2overview}
\end{figure*}

\textbf{Evaluation---imaging gastruloids:} Using flanged tubing and luer lock connections, we successfully aspirated single gastruloids into the FEP tube without trapping air bubbles. The gastruloid rests on a flat steel pin, which does not block the incident light sheet; see Figure \ref{fig:Gen2overview}(h), as is the case when the soft, synthetic polymer seals the tube. These improvements made it possible to identify single cells in a 3-day-old gastruloid (diameter: 0.3 mm), as shown in Figure \ref{fig:Gen2overview}(i,j), imaged with a camera exposure time of 1000 ms and a laser power of 20 mW. 

\textbf{Evaluation---Imaging cellulose nanocrystals:}
The aspiration and sealing method also enabled us to image untagged cellulose nanocrystals (CNCs; a type of liquid crystal) with light sheet microscopy for the first time by utilizing their autofluorescence \cite{johns2022autofluorescence,kalita2015isolation} in blue laser light; see Figure \ref{fig:Gen2overview}(k,l). To maximize the inherently weak autofluorescence signal, we removed the neutral density (ND) filter, set the laser power to as high as 60 mW out of the cavity, and used an exposure time of 2000 ms. 

Figure \ref{fig:Gen2overview}(k) is a single 2D cross section showing mesoscopic 'banded' structures of the cholesteric phase known as 'fingerprint textures'\cite{gray2015chiral} (CNC concentration: 4 wt$\%$, probe sonicated for 30 minutes prior to imaging, and aspirated from the bottom phase of a biphasic CNC suspension.). 
In contrast to polarized light imaging (PLI), which is traditionally used to visualize liquid crystals, light-sheet microscopy offers true optical sectioning because it excludes signal from out-of-focus material. This means we can determine the distance between lines in the fingerprint texture (corresponding to half the pitch of the chiral nematic structure) and their widths more accurately, with typical values of 8-9 $\mu$m and 5 $\mu$m in our experiments. The optical sectioning capability also allows resolution of topological defects, which is difficult in PLI where they typically appear as dark spots \cite{dunmur2010soap}. Because the SPIM signal comes from autofluorescence, the local concentration (and its gradients) can in principle be determined, complementing PLI measurements, where the optical signal comes from the orientation of the CNCs.

Figure \ref{fig:Gen2overview}(l) is a 3D image of CNC mesoscale structures formed from an image stack of 195 optical cross sections. From these 3D data, it is possible to determine how the pitch and width of individual bands vary with position in 3D, and to follow individual defects to backtrack the history of coalescing tactoids that formed the cholesteric phase. In addition, loading the suspensions by syringe-and-needle aspiration into the tube mimics extrusion-based 3D printing and paves the way for controlled studies of how local shear rates influence CNC mesoscale morphology, with potential applications in bioprinting and the development of new materials. 

\textit{Drift:} By tracking three individual defects, we found that the suspension drifts downward at a mean rate of 13 $\mu$m (std: 5 $\mu$m) during a typical 3D imaging experiment (duration: 390 seconds). This drift is comparable to the distance between bands (8-9 $\mu$m), but minimal compared to the tube radius (0.8 mm), so that the system can be treated as quasi-static. The drift is due to CNCs sedimenting (as they are heavier than the solvent, water), incomplete tube sealing, or a combination of both. Reducing exposure time and the number of optical sections can reduce drift during a 3D imaging experiment. 

\begin{comment}
The observed mesoscopic 'banded' structures are known as 'fingerprint textures' in polarized light imaging of CNC suspensions and films. We note that, in polarized light, the texture results from birefringence and provides a 2D interrogation of the structure.
\end{comment}

\section{Generation 3: Multi-channel imaging}
\label{sec:generation3}
\subsection{Prototype 3A: New laser with 4 channels}
While a single laser enabled us to image live and fixed specimens, our setup needs to support imaging multiple fluorophores, bringing us closer to our goal of visualizing different germ layers and how they evolve (differentiate, proliferate, and migrate) in live gastruloids. This advancement would also enable the imaging of different fluorescent markers in other biological specimens, such as zebrafish and Drosophila, dramatically enhancing the versatility of our setup.

\textbf{Implementing the new laser:} To facilitate imaging of different fluorophores, we replaced our blue ($\lambda$=488nm) laser with a new laser (Cobolt 6 series, Hübner Photonics) with "all four" wavelengths, that is, ultraviolet (UV; $\lambda$=375 nm, maximal power: 70 mW, beam diameter at aperture: 700$\pm$100 $\mu$m; product no.: 0375-06-01-0070-100-C), green ($\lambda$=561 nm, maximal power: 50 mW, beam diameter at aperture: 700$\pm$70 $\mu$m; product no.: 0561-06-91-0050-100-C), red ($\lambda$=647 nm, maximal power: 130 mW, beam diameter at aperture: 700$\pm$100 $\mu$m; product no.: 0647-06-01-0130-100-C), as well as blue ($\lambda$=488nm, maximal power: 60 mW, beam diameter at aperture: 700$\pm$100 $\mu$m; product no.: 0488-06-01-0060-100-C), corresponding to DAPI, red fluorescent protein (RFP) or mCherry, Alexa Fluor647, and green fluorescent protein (GFP) fluorescent markers. The UV, blue, and red lasers are modulated laser diodes (MLDs), while the green laser is a diode-pumped laser (DPL). A laser combiner (C-FLEX, Hübner Photonics; product no.: C6-11-00-00-100) combines the lasers and interfaces with the computer via USB, and its power and current are controlled using the manufacturer's Cobolt Monitor software. To minimize light scattering from the sample and the tube, an emission filter (OD6 ULTRA Quad-Bandpass, Alluxa) with four bands (379 to 401 nm, 463 to 485 nm, 538 to 556 nm, 627 to 641 nm) that block the incident wavelengths is placed in one of the two slots on the infinity space tube (self-made; for CAD drawings, see Ref.\cite{webpagekeyOpenSPIM_parts}), in front of the tube lens (U-TLU, Olympus; f=180 mm, M=1$\times$) and camera, replacing the emission filter (Chroma ZET 488/561m) used previously.

\textbf{Evaluation:} The setup enables imaging of multiple fluorophores, but switching between laser wavelengths takes several seconds, making live imaging impractical. 
%The 561 nm laser is particularly slow since it is a diode pumped laser (DPL), while the other lasers are quicker  changing between the different wavelengths of the laser is  

%From Dag email 20250114: "The 561 laser is a Diode Pumped Laser (DPL) whereas the others are modulated laser diodes (MLD). It is correct that the DPL is 10 000 times slower than the MLDs, but the modulation frequency is still 1 kHz! That means that in 1 millisecond you can have the intensity you want. The only reason you may have to wait for a second or more is that you turn the laser on and off (which I doubt is healthy for the laser)." 

%\subsubsection{Conclude} We need to switch between the different lasers more quickly. Also, to enable automated multi-view imaging, the laser must be controlled by Micro-Manager\cite{edelstein2010computer}. 

\subsection{Prototype 3B: Arduino microcontroller for fast switching between lasers}

\begin{figure*}[p]
    \centering
    \includegraphics[width=0.8\linewidth]{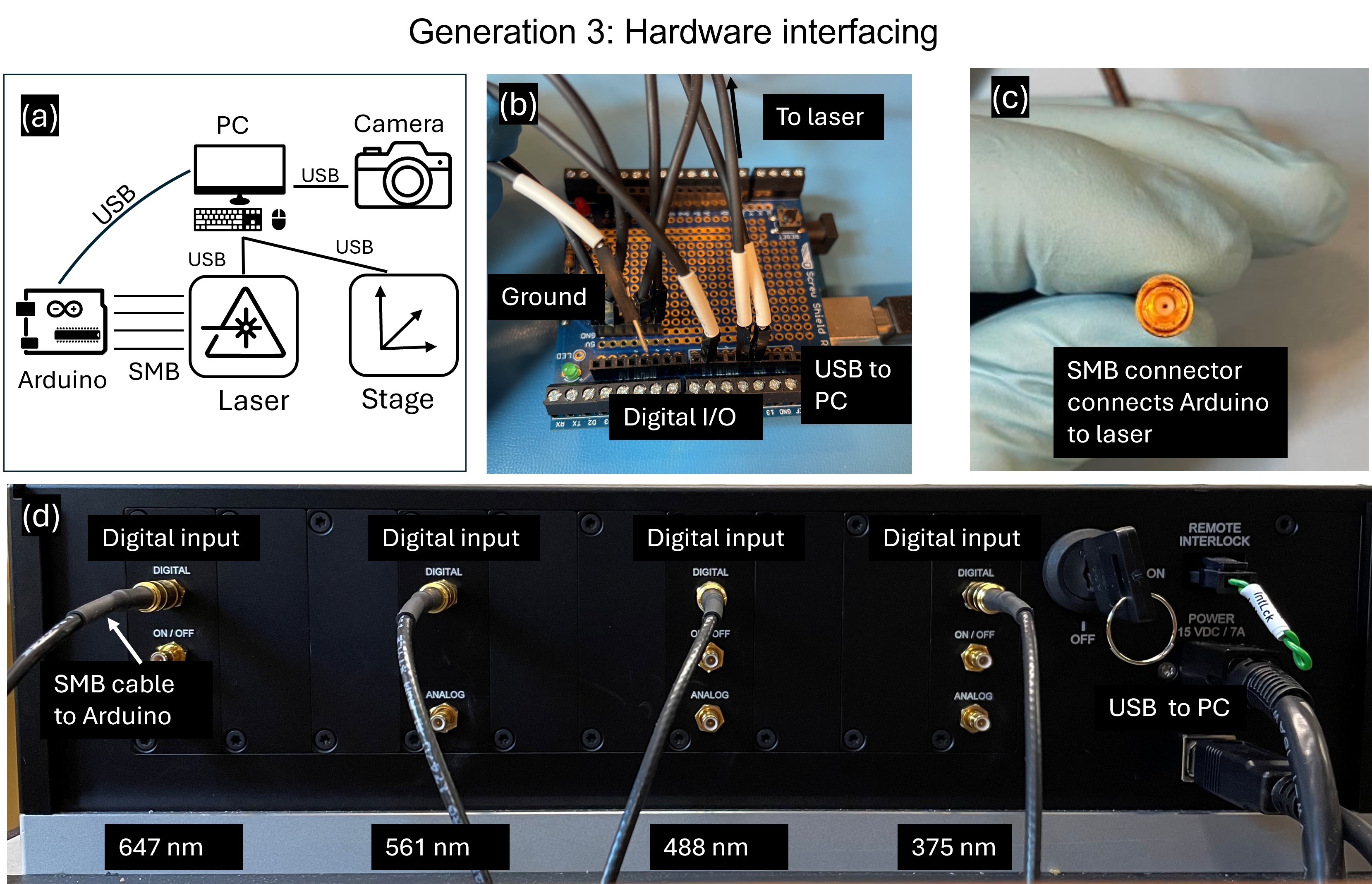}
    \caption{Hardware interfacing in Prototype 3B of Generation 3. \textbf{(a,b)} The camera, microscope stage, laser, and Arduino microcontroller are connected to the computer via USB cables and are controlled by Micro-Manager. To switch quickly between the different laser channels (UV, blue, green, and red), the Arduino microcontroller sends trigger pulses via SMB cables. \textbf{(c,d)} The SMB cables plug into digital input ports on the laser via SMB connectors, and a USB cable interfaces the laser with the computer.}
    \label{fig:Gen3interfacing}
\end{figure*}

To switch quickly between wavelengths, the laser can be modulated with a digital TTL (Transistor-Transistor-Logic) square-wave signal. An Arduino microcontroller (Arduino Uno Rev3 SMD, Arduino) transmits this digital signal to the four lasers via digital SMB (Subminiature B) cables (095-700-37-M200, Amphenol RF) and ports, and to the computer via USB; see Figure \ref{fig:Gen3interfacing}. The Arduino microcontroller, stage, camera, and laser are all integrated into Micro-Manager on the computer to enable fully automated multi-view operation. For instructions on integrating the Cobolt laser and Arduino microcontroller in Micro-Manager, see Refs. \cite{micromanagerCoboltOfficial,micromanagerArduino}.

To connect the laser's SMB ports (Figure \ref{fig:Gen3interfacing}) to the Arduino microcontroller headers, we cut the SMB cables in half because they had SMB connectors on both ends. The cables are coaxial, with a central conductor surrounded by a braided copper shield. We crimped thin wires to both the central conductor and the shield. At the other end, we crimped on DuPont connectors. We then applied heat-shrink tubing to cover the exposed conductors.

The central conductors were connected to digital outputs on the Arduino, and the shields were connected to Ground (GND); see Figure \ref{fig:Gen3interfacing}(b). On the Arduino, we mounted a Screwshield (Proto-Screwshield (Wingshield) R3 Kit for Arduino, Adafruit) to provide additional GND connection points and enable more robust wiring.

%Initially, we were not able to modulate the laser using the Arduino microcontroller. We hypothesized that this was due to a problem with the laser, either because the SMB ports on the laser were not functional or due to an erroneous setting in the laser software. To test our hypothesis, we disconnected the Arduino from the laser and instead used a function generator to send a square-wave signal via the laser's SMB ports. As this enabled us to modulate between different wavelengths, we could rule out the laser as being the source of the error. We then used the function generator to test the SMB ports and cables, as well as the Arduino microcontroller, and found them all to be functional. As such, we concluded that we were unable to modulate the laser with the Arduino due to an interfacing problem with the computer. After some trial and error, we found that the Arduino was not properly configured in Micro-Manager. Once we solved this issue,

%Fig. 9

\begin{figure}[t]
    \centering
    \includegraphics[width=0.5\linewidth]{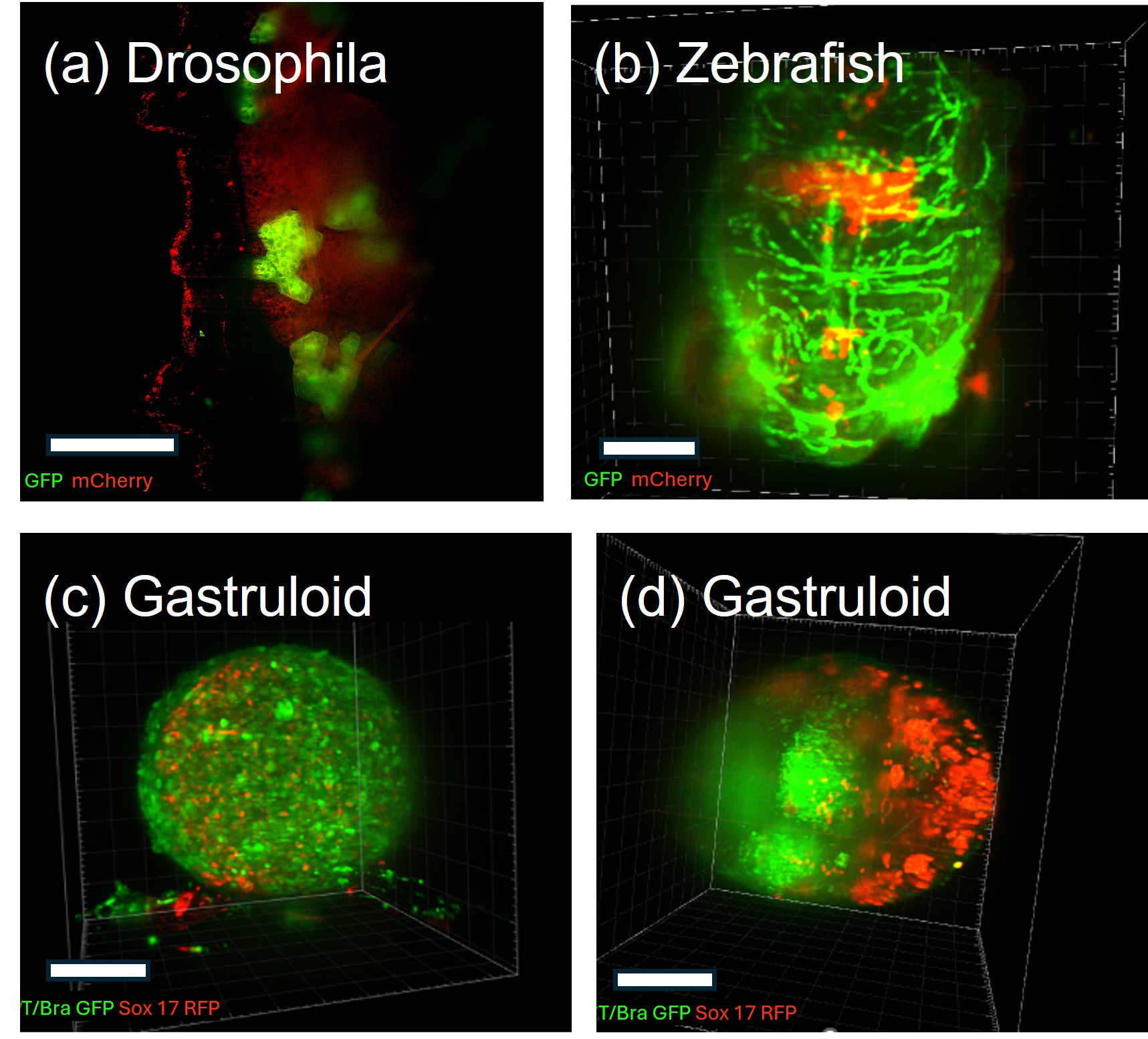}
    \caption{Multichannel imaging in Generation 3: By illumination by the blue and green lasers (green and red emission), we image \textbf{(a)} the fat body tissue in Drosophila larvae, where genetically manipulated cells are expressing GFP, while all cells express 3xmCherry-Atg8a under control of the Atg8a promoter,  \textbf{(b)} mCherry tagged cancer cells invading the GFP-tagged blood vessels in the brain of a zebrafish larva (6 days post fertilization, dpf), \textbf{(c)} maximum intensity projections of mouse gastruloids at 3 days post-aggregation (dpa) and at \textbf{(d)} 4 dpa. In (c) and (d), marker genes are labeled with Brachyury-GFP and Sox17-RFP reporters, corresponding to the mesoderm (green) and endoderm (red) germ layers, respectively. Panel (a) shows a single (2D) optical cross-section, while the other panels show 3D-rendered data obtained from 2D image stacks. The image in panel (a) is deconvolved using Generic Deconvolution, and the images in panels (c) and (d) are deconvolved using Huygens SVI. The image in panel (b) is not deconvolved. Scale bars: 100$\mu$m.}
    \label{fig:Gen3imaging}
\end{figure}

\textbf{Evaluation:} 
By integrating the stage, laser, Arduino, and camera in Micro-Manager, we could switch between lasers in milliseconds and achieve fully automated multi-view, multicolor 3D imaging of fixed biological specimens; see Figure \ref{fig:Gen3imaging}. Panel (a) is a single optical section 90 $\mu$m inside a Drosophila larva (exposure time: 100 ms, z-range: 397 $\mu$m; power setting for the blue (488 nm) laser: 10 mW, power setting for the green (561 nm) laser: 30 mW). Panel (b) is a 3D rendered image showing mCherry (red) tagged cancer cells invading the GFP-tagged blood vessels in the brain of a zebrafish larva (power (both green and blue lasers): 20 mW, exposure time: 200 ms, z-range: 530 $\mu$m, z steps: 1.542 $\mu$m). Panels (c) and (d) are maximum intensity projections of mouse gastruloids at 3 and 4 days post-aggregation (dpa; power (both green and blue lasers): 20 mW, exposure time: 200 ms, z-range: 400 $\mu$m, z steps: 1.542 $\mu$m). In (c) and (d), marker genes are labeled with Brachyury-GFP and Sox17-RFP reporters, corresponding to the mesoderm and endoderm germ layers.  

To achieve our goal of performing live imaging of tissue cultures for several hours, we need to control environmental conditions. Building an environmental chamber for such cultures is the aim of the last two generations of our setup. We start by building a temperature control unit.    

\begin{comment}

\subsection{Setup}
\begin{itemize}
    \item New laser with 4 wavelengths (375, 488, 561, and 647nm)
    \item (a) Laser controlled with Hubner software
    \item (b) Control laser with Arduino
    \item New quad bandpass emission filter
\end{itemize}

\subsection{Postprocessing}
\begin{itemize}
    \item Merge channels in ImageJ
\end{itemize}

\subsection{Utilization}
\begin{itemize}
    \item Image cell types with different fluorescent markers in fixed samples (gastruloids, zebrafish and drosophila). 
    \item We visualize mesoderm and endoderm germ layers in gastruloids, cancer cell distribution and clustering in zebrafish brains, different cell types in drosophila.  
\end{itemize}

\end{comment}
\section{Generation 4: Temperature control}
\label{sec:generation4}
\textbf{Design criteria:}
The temperature control unit must meet the following criteria. First, the temperature control unit should not induce thermal drift of optical components in the illumination or detection paths. Second, the temperature in the viewing chamber should be uniform so the sample experiences the same temperature regardless of its position in the chamber, since the temperature is typically not monitored during imaging. Third, the temperature should equilibrate quickly to the operator-set temperature to enable fast, reliable experiments. 

We considered three different strategies for controlling the temperature. The first is to enclose the entire setup in an environmental chamber and circulate heated air to maintain the desired temperature. However, we quickly abandoned this idea because it is prone to thermal drift in the mirrors and lenses along the illumination path unless the temperature is kept constant, which is impractical in our case. As such, it is better to localize the temperature control unit within the viewing chamber, which leaves us with two options. The first option is to use a thermoelectric (Peltier) plate. However, these devices can be difficult to set up, can easily overheat, and, because they dissipate heat only from below, large temperature differences should be expected in large viewing chambers like those used in light-sheet microscopes. Given these limitations, we decided to control the viewing-chamber temperature by circulating heated water instead.      

\subsection{Prototype 4A: Pumping water at physiological temperature into the viewing chamber}

\textbf{Building:}
Figure \ref{fig:Gen4overview}(a) shows the initial design. Flowing hot water at physiological temperature, mimicking in vivo conditions (e.g., 37 $^\circ$C for mammalian tissue culture such as gastruloids), is pumped directly into the viewing chamber by a circulator pump (GR 150, Grant Instruments), which connects to the viewing chamber via silicone tubing. Nickel nipples (inner diameter: 2.2 mm) are inserted into holes drilled through the acrylic viewing chamber to interface the silicone tubing with the chamber. 

\textbf{Evaluation:}
The pump's high flow rate, which cannot be adjusted, caused water to overflow the viewing chamber. %Clamping the tubing could reduce the flow rate; however, this causes bubbles to form in the viewing chamber, severely degrading image acquisition.  

%Figure 10 here

\subsection{Prototype 4B: Heating channels in the acrylic viewing chamber prevent overflow}
In Prototype 4B, heated water from the pump instead flows through channels (diameter: 4\,mm) drilled through the two side walls of the acrylic viewing chamber, not holding the objectives; see Figure \ref{fig:Gen4overview}(b). This circumvented the overflow issue, but due to poor heat transfer through the walls of the acrylic chamber, the temperature varied by up to 10$^\circ$C from top to bottom. As such, we refined this design to achieve a more uniform temperature in the chamber.   

\subsection{Prototype 4C: Aluminum viewing chamber ensures uniform temperature}
\begin{figure}[p]
    \centering
    \includegraphics[width=0.45\linewidth]{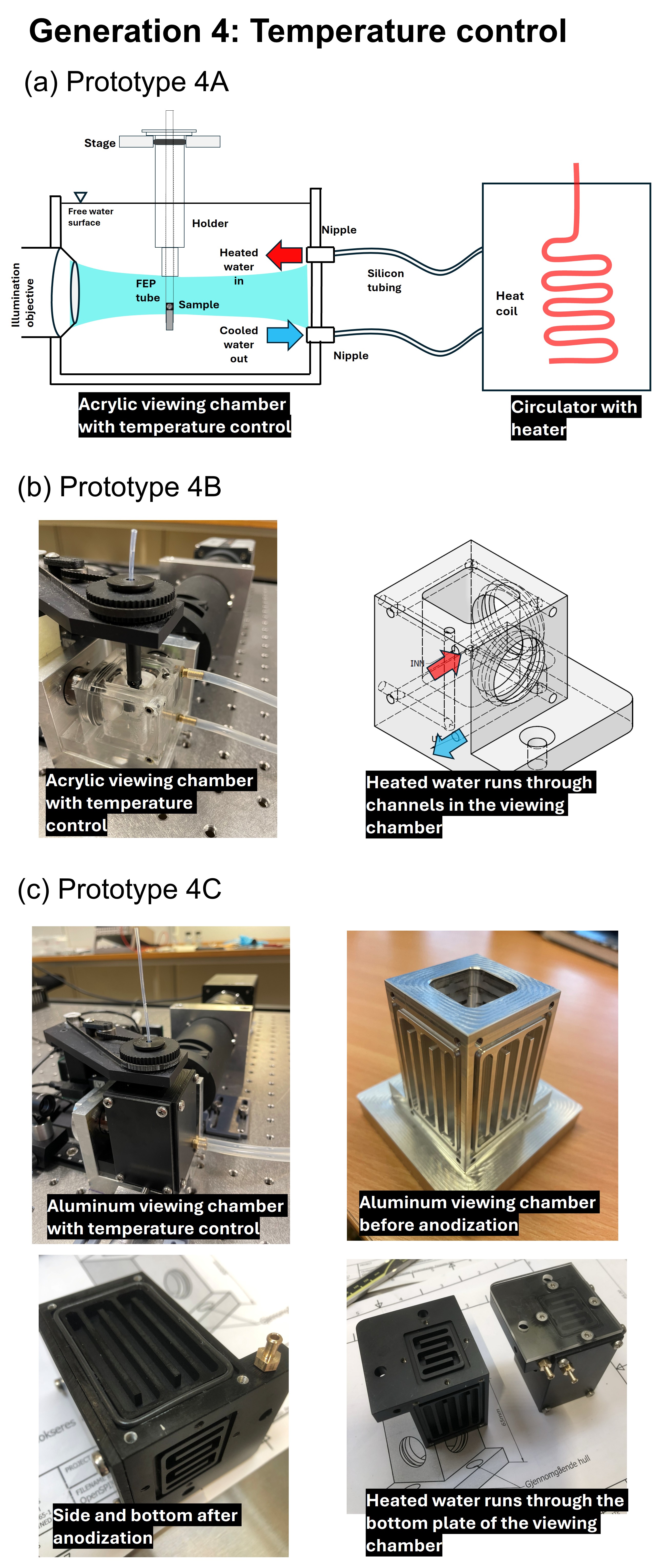}
    \caption{Different iterations of a temperature control unit in Generation 4 of the setup. \textbf{(a)} Prototype 4A: A circulator pumps heated water directly into the viewing chamber. This provides a uniform temperature in the viewing chamber, but the strong pump flow easily causes the water to overflow. \textbf{(b)} Prototype 4B: Heated water from the circulator flows through drilled channels in the acrylic viewing chamber walls. This circumvents the overflow issue, but because heat transfer through the acrylic chamber is poor, the temperature varies by 10 $^\circ$\text{C} from top to bottom. \textbf{(c)} Prototype 4C: Replacing the viewing chamber material with aluminum dramatically improves heat transfer. The milled zigzag channels further increase heat transfer through the block, resulting in a uniform temperature throughout the entire viewing chamber.}
    \label{fig:Gen4overview}
\end{figure}

In Prototype 4C, three main steps were taken to enhance the heat transfer; see Figure \ref{fig:Gen4overview}(c). First, the viewing chamber was made of aluminum rather than acrylic because aluminum conducts heat much more efficiently (heat transfer coefficient of acrylic: 20 $\text{W}/(\text{m}\cdot{}^\circ\text{C})
$\cite{engineeringtoolbox_thermal_conductivity}; heat transfer coefficient of aluminum: 164 $\text{W}/(\text{m}\cdot{}^\circ\text{C})
$\cite{engineeringtoolbox_thermal_conductivity_metals}). 
Second, the heat transfer surface area was increased by milling 4 by 5 mm zigzag channels in the two side walls of the viewing chamber, not holding the objectives (see the top right panel in Figure \ref{fig:Gen4overview}(c)), as well as 3 by 5 mm zigzag channels in its bottom surface (not shown). After milling, we anodized the chamber black to eliminate toxic effects on cells and minimize autofluorescence and light scattering. The side-wall channels are covered with a 3 mm-thick aluminum plate, sealed with O-rings to prevent leakage, and the inlets and outlets are connected to the pump via silicone tubing interfaced with tube nipples (inner diameter: 2.2 mm). The third and final step to ensure a uniform temperature was to mount the viewing chamber on a 3 mm-thick acrylic plate (bottom panels in Figure \ref{fig:Gen4overview}(c)) to provide thermal insulation from the optical table, which otherwise acts as a heat sink.

\textbf{Evaluation:}
The enhanced heat transfer and insulation from the optical table provided a uniform temperature throughout the viewing chamber, which ramped up to 37 $^\circ$C in six minutes, as measured with a thermocouple (Standard ST-612, Clas Ohlson). 

\section{Generation 5: Flow control}
\label{sec:generation5}

\begin{figure*}[]
    \centering
    \includegraphics[width=0.8\linewidth]{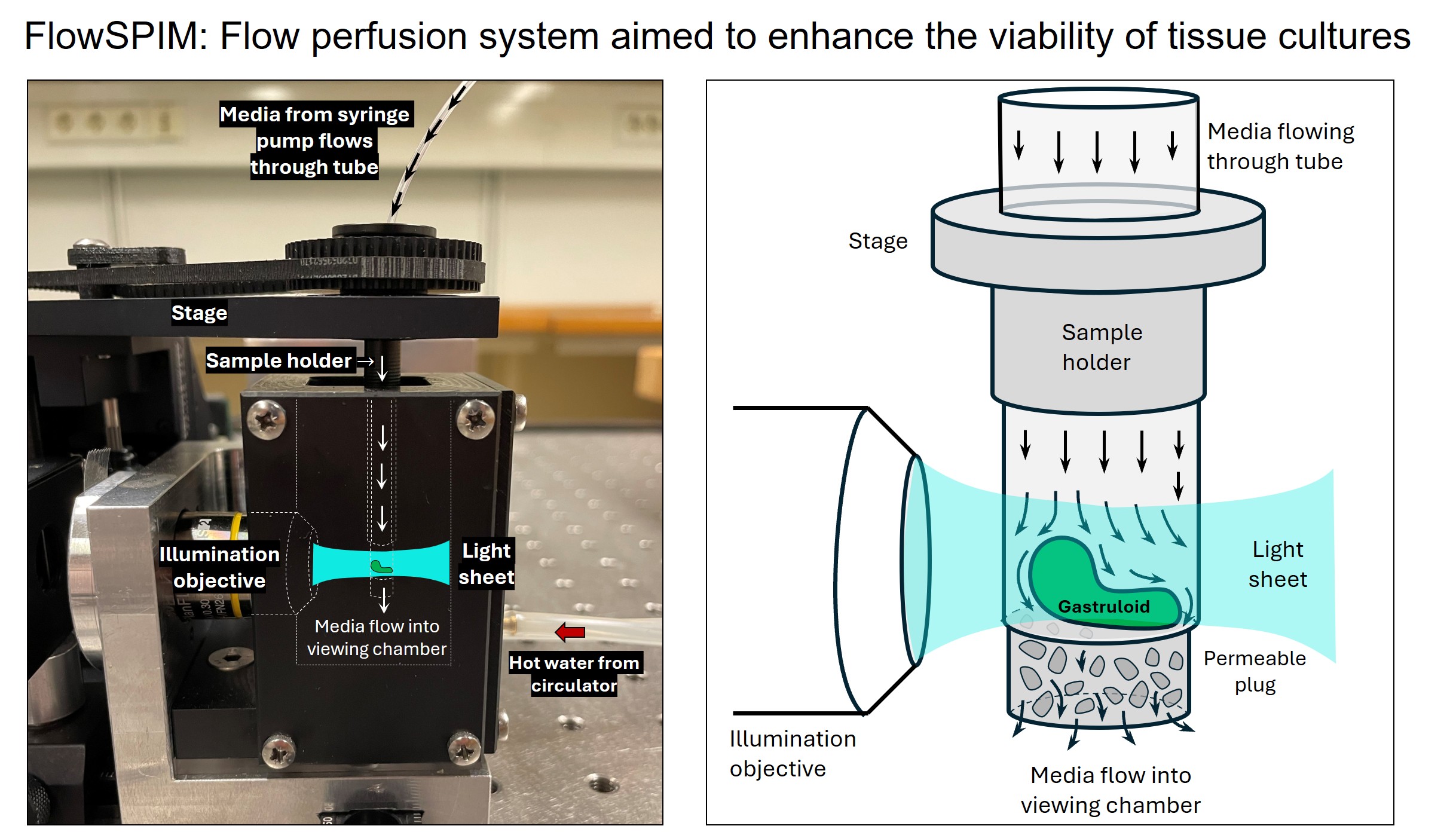}
    \caption{FlowSPIM: Overview of flow perfusion system proposed in Generation 5 aimed to enhance the viability of 3D tissue cultures such as gastruloids. A syringe pump delivers flowing media through the capillary tube 
    at a physiologically relevant flow rate to provide the cells with fresh oxygen and nutrients, and to remove excess \text{CO\textsubscript{2}} to maintain favorable \text{pH}-levels. The tissue culture rests on a porous plug, and the cell media flows directly into the viewing chamber. In Prototype 5A, the porous plug was made by extracting a cylindrical section from a pipette tip using a biopsy punch, and in Prototype 5B and 5C, a porous steel plug was used instead.}
    \label{fig:flowperfusion}
\end{figure*}

\textbf{Designing the flow control unit:}
To keep tissue culture viable over time, our incubation chamber must mimic in vivo conditions. As such, we need to provide physiological pH, oxygen, and shear stress levels, in addition to physiological temperature. To this end, we propose to continuously exchange the media by flowing it through the capillary tube, inspired by perfusion systems used in organ-on-a-chip devices\cite{huh20113d,huh2010reconstituting,kim2012humanGutChip,zhang2016heartOnChip,bauwens2014liverOnChip,eschenhagen2017engineeredHeart}. Figure \ref{fig:flowperfusion} shows the working principle of the proposed flow perfusion system.  

The flow perfusion system must meet the following design criteria. First, to enable flow through the tube, the plug sealing it from below needs to be permeable. Second, to maintain a constant \text{pH} level, the flow system needs to be gas-tight to prevent \text{CO\textsubscript{2}} leakage, since the \text{pH} level is a function of the amount of \text{CO\textsubscript{2}} dissolved in the cell media. Third, to ensure the flowing media reaches physiological temperature by the time it reaches the sample, heat must be efficiently conducted from the (heated) water in the viewing chamber to the tube. 

\subsection{Prototype 5A: The initial flow control unit}
\textbf{Building the flow control unit:}
Figure \ref{fig:flowperfusion} is a schematic of the flow setup, and Figure \ref{fig:holders}(c) is an image of Prototype A of the sample holder. A syringe pump (Legato 100, KD Scientific) provides flow by pushing the plunger of a Luer-lock syringe filled with N2B27 cell media (see the Appendix for the protocol) to deliver flow velocities of typically 10 $\mu$m/second. The syringe is connected to the flanged capillary tube via an airtight Luer lock, and the capillary tube is sealed at the other end with a porous Teflon plug, allowing perfusion flow into the viewing chamber. We made this plug by extracting a cylindrical section from a pipette tip using a biopsy punch.  Finally, temperature equilibration channels were made by drilling neighboring holes through the lower part of the micromachined holder (same holder as used in previous generations of the setup) to bring the tube into direct contact with the heated water in the viewing chamber to facilitate efficient heat transfer to the flowing media in the capillary tube. 

%Figure 10 here

\textbf{Evaluation of Prototype 5A:} First, we checked for leaks and air bubbles in the flow system; both were absent at flow rates up to 100 $\mu$L/min. We then evaluated the media pH. To this end, we filled the syringe a capillary tube with DMEM containing phenol red, which changes color in response to changes in pH and monitored the color. During a 24-hour flow experiment (1 µL/min), we observed no color change, indicating a stable pH maintained by the airtight Luer lock connection.   

However, the setup has several limitations. First, because of the plug's rough surface, we found it hard to insert into the capillary tube. Second, the tube holder slipped in the tube because it could not withstand the torque during sample rotation. Third, the temperature equilibration channels were limited to a very small area at the lower end of the tube holder, as drilling more holes would cause it to break.

\textbf{Conclusion:}
To address these limitations, we need a new holder with longer temperature equilibration channels to ensure the flowing media are heated sufficiently to reach physiological temperature by the time they reach the sample at the end of the tube. The new holder must also connect tightly to the stage to ensure slip-free rotation. Finally, we need a new plug that is easier to push into the tube and fits more snugly. 

\subsection{Prototype 5B: 3D printed holder ensures efficient heat transfer to the media}
\begin{figure*}[t]
    \centering
    \includegraphics[width=0.8\linewidth]{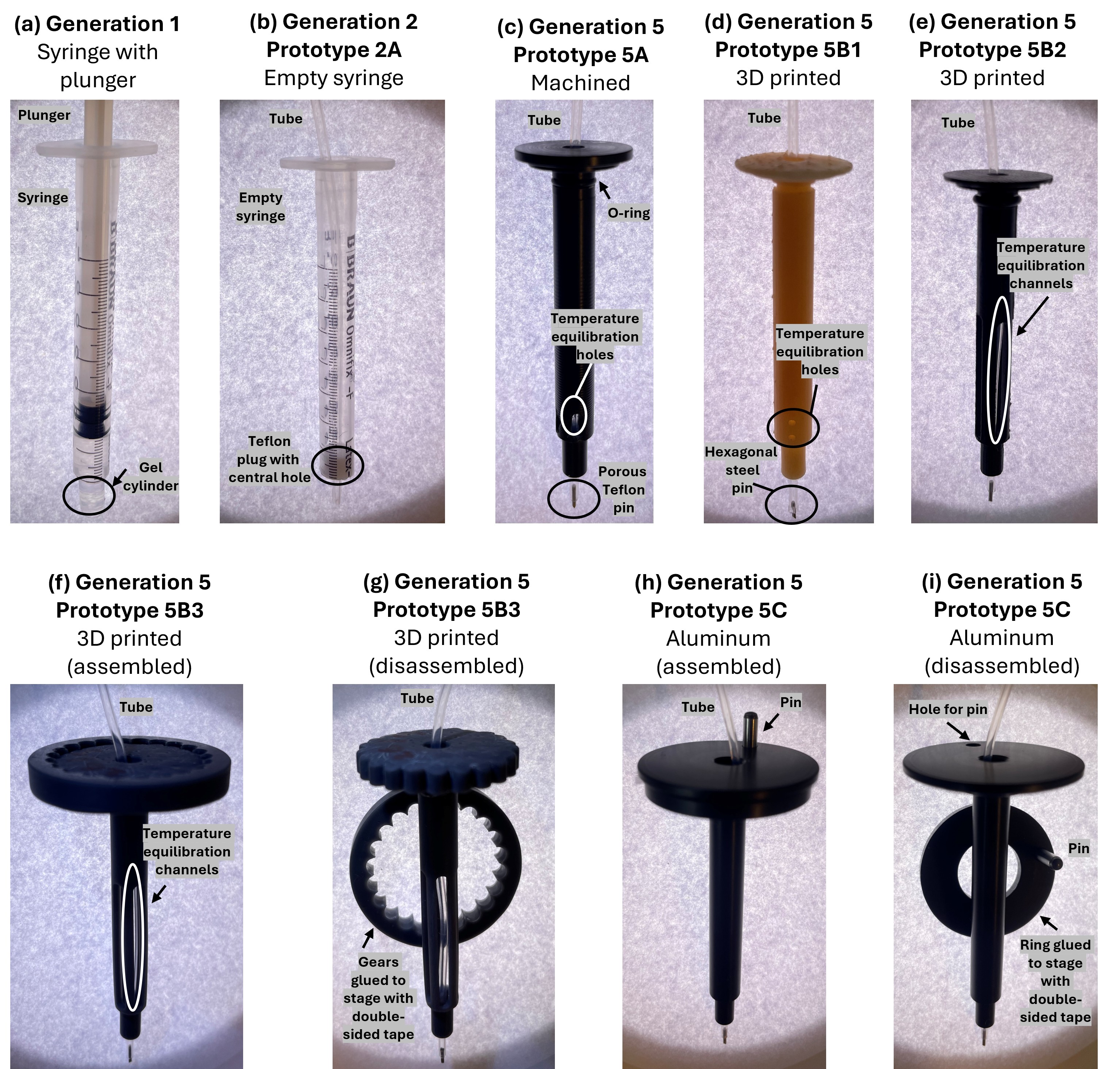}
    \caption{Overview of sample holders. \textbf{(a)} Generation 1: Gel-based sample mounting using a syringe. \textbf{(b)} Generation 2 Prototype 2A: Sample is mounted in a capillary tube held in place by a Teflon plug inside a syringe. \textbf{(c)} Generation 5 Prototype 5A: Machined holder with drilled temperature equilibration channels to facilitate heat transfer to the sample mounted inside a capillary tube. The same holder was used in Generation 2 Prototype 2B, and in Generations 3 and 4, but without the temperature equilibration channels. \textbf{(d)} Generation 5 Prototype 5B1: 3D printed holder with temperature equilibration holes. \textbf{(e)} Generation 5 Prototype 5B2: 3D printed holder with temperature equilibration channels along the side. \textbf{(f,g)} Generation 5 Prototype 5B3: 3D-printed holder with temperature equilibration channels and gears to prevent slipping on the stage during rotation. The gears are glued around the sample holder hole on the stage. \textbf{(h,i)} Generation 5 Prototype 5C: Machined aluminum holder. The pin is threaded through a hole in the holder to prevent it from slipping on the stage during sample rotation. The gears are glued around the sample holder hole on the stage.}
    \label{fig:holders}
\end{figure*}

To address these issues, we turned to 3D printing. This let us increase the wetted area between the tube and the water in the viewing chamber by printing temperature-equilibration channels along the holder's entire length, enhancing heat transfer. Through 3D printing, we also made a much tighter connection between the stage and the holder using a gear interlock system to prevent the holder from slipping in the stage during rotation. Finally, we replaced the porous Teflon plug with a stainless steel pin with a hexagonal cross-section to allow flow perfusion; see Figure \ref{fig:holders}(d).

\textbf{Building and evaluating the 3D printed sample holders:}
We made three holder prototypes, as described below. 

\textit{Prototype 5B1:} To check for tolerances, we first made a 3D-printed prototype in PLA plastic (Polylactic Acid); see Figure \ref{fig:holders}(d), based on the same CAD drawings used to make the micromachined holder in Prototype 5A (Figure \ref{fig:holders}(c)). We found the outer dimensions of the 3D print to match those of the micromachined holder, but the hole running through the length of the holder to be too narrow. To increase the hole radius, we used a Dremel tool, which allowed us to thread the tube through the hole and to make a tight connection. 

%Figure 12 around here

\textit{Prototype 5B2:} To improve heat transfer to the tube, we increased the tube's wetted area by adding temperature-equilibration channels along the holder; see Figure \ref{fig:holders}(e). As in Prototype 5B1, we also made Prototype 5B2 in PLA. Unfortunately, the combination of brittle PLA and the temperature-equilibration channels caused the holder to break easily. In addition, the holder slipped in the stage during sample rotation. 

\textit{Prototype 5B3:} To enhance the mechanical stability, Prototype 5B3 was made in Acrylonitrile butadiene styrene (ABS), which is more ductile than PLA; see Figure \ref{fig:holders}(f,g). This new and improved holder also has a gear interlocking system that efficiently suppresses slipping during sample rotation; see Figure \ref{fig:holders}(g). However, during long experiments, the holder bent due to insufficient mechanical stability, causing sample drift during imaging.  

\subsection{Prototype 5C: Aluminum sample holder enhances heat transfer and minimizes sample drift}
\textbf{Design:}
In previous iterations, temperature equilibration channels were needed to facilitate heat transfer to the tube, since plastics are poor heat conductors. As such, if the holder were made of a more heat-conductive material, it would not need temperature equilibration channels, making it sturdier and less prone to breaking. Aluminum is a good heat conductor, far better than plastics, and, with better mechanical and thermal stability, it should not bend during long-time-lapse imaging, thereby efficiently reducing sample drift. As before, the holder must interlock with the stage to ensure slip-free rotation during multi-view imaging.  

\textbf{Building:}
Figure \ref{fig:holders}(h,i) shows Prototype 5C of the sample holder. The aluminum holder was anodized black to minimize light scattering and improve biocompatibility. The holder interlocks with the stage via a ring glued to the microscope stage with double-sided tape. A pin on this ring is threaded through a hole in the holder to prevent it from slipping on the stage during sample rotation. 

\textbf{Evaluation:} The aluminum holder interlocks well with the stage and ensures mechanical stability without thermal drift during long experiments. 

\subsection{Prototype 5D: FlowSPIM - Live imaging of gastruloids}
\textbf{Experimental design:} To evaluate the flow incubation system, we performed live imaging of a mouse gastruloid at 4 days post-aggregation (4 dpa) in the blue and green laser channels, corresponding to the mesoderm and endoderm germ layers marked with Brachyury-GFP and Sox17-RFP reporters, respectively. We acquired a z-stack of images every 30 minutes, matching typical cell-migration timescales as determined by preliminary confocal imaging experiments. The z-step size (between successive optical slices) was 10 $\mu$m, sufficient to resolve cell shape \cite{ho2025spontaneous} while minimizing phototoxic dose. To further minimize phototoxicity, the imaging was performed from only one side, and the laser power was set as low as 4\text{mW} out of the cavity, with an approximate power of 0.5\text{mW} in the imaging chamber, as calculated from the ratio of the frontal area of the beam after and before the slit aperture. To obtain sufficient optical contrast to distinguish neighboring cells, we set the camera exposure time to 1000 and 500 milliseconds for imaging in the blue and green channels, respectively. We needed such long exposure times because the samples are thick, dense objects without clearing, and the Brachyury-GFP reporter signal is particularly hard to visualize in the blue channel due to weak, transient expression and diffuse localization. The experiments were performed at 37 $^\circ$C in the viewing chamber, as measured with a thermocouple, and the flow velocity of the cell media through the tube was typically 10 $\mu$m/s.   

\textbf{Experimental results:} 
Figure \ref{fig:livegastruloid_decon} is an optical cross-section of a live mouse gastruloid at 4 dpa, 30 minutes after mounting in the experimental incubation chamber. Here, the T/Bra GFP (green) signal indicates cells in the mesoderm germ layer, and the Sox17 RFP (red) signal indicates cells in the endoderm germ layer, with individual nuclei visible. 

This simple imaging experiment demonstrates live imaging of mouse gastruloids with the FlowSPIM imaging platform (Generation 5 of the setup). These fragile samples were viable for about one hour.

%Unfortunately, we were not able to keep these fragile samples viable for more than about one hour, and we were not able to identify cell migration or shape evolution of the overall gastruloid. We hypothesize that the lack of cell viability is due to toxins released by the stainless steel pin sealing the capillary tube, and we are currently developing a new mounting technique to address this issue.   

%Figure 13 aound here

\section{Discussion: Routes of improvement and contribution to the field}
\label{sec:discussion}
%\textcolor{blue}{This section was added after the initial submission and attempts to address the referees' comments below (in red).}

%\textcolor{red}{Ref1: In the conclusion sections, please comment on what can be improved further, based on image quality and acquisition speed. Resolution? Removal of autofluorescence and stray light? Which PSFs should be used for deconvolution? How to minimize striping effects observed in Fig. 13? }

%\textcolor{red}{Ref2. 1. The authors themselves write in their introduction that the article should serve as a guide for other groups building their own similar setup. However, the rest of the article (while well written) describes their own work, and potential readers are presumably expected to draw their own conclusions. To make the troubleshooting lessons more applicable to the community, the authors should add a section with explicit troubleshooting instructions, e.g., if a group wanted to retrofit a similar FEP tube live-imaging chamber into an existing SPIM.}

%\textcolor{red}{Ref2. 3. Live-imaging organoids (though not necessarily gastruloids) have been demonstrated a few times before, with different approaches and tradeoffs. The authors should discuss at least the following articles and what their own work brings to the table:}

%\textcolor{red}{Ref2: 4b. There should be a discussion section in which the authors set their own work in the context of the larger light-sheet imaging field and/or live-imaging field (see point 3 above).}

\textbf{Recommended practices:}
This section presents general recommendations that can guide others who want to build their own SPIM. Table \ref{tab:BOM_FlowSPIM} lists the bill of materials for the FlowSPIM (Prototype 5D), the last prototype in the setup. 

\textit{Generation 1 - Optical setup:} 
In our experience, the most difficult part of building a microscope based on the OpenSPIM platform was aligning the laser, and we optimized the final optical configuration (Prototype 1D) to ease this task. Especially, the two alignment mirrors were placed at the focal points of the first two lenses to enable pure adjustment of the light sheet angle relative to the detection focal plane with the first mirror (without changing the position), and pure adjustment of the light sheet position along the optical (z) axis with the second mirror (without changing the angle). To further ease laser alignment, the 1/2$^{\text{''}}$ spherical lenses could be replaced with 1$^{\text{''}}$ optics, since the latter is less prone to spherical aberrations. To keep the spherical lenses perpendicular to the laser beam, they could be mounted on a cage system, replacing the dovetails used in this work. 

%To obtain a good measurement of the point spread function (PSF), the beads should be smaller than the lateral resolution of the optical system, given by the Rayleigh criterion: $d_{beads} <\text{0.61}\lambda_{laser}/\text{NA}$, which places an upper bound on the particle size. To further improve the PSF calculation, reduce the particle size below this limit (0.6 $\mu$m in our setup), since the particles themselves contribute to the measurement. To fully quantify the optical system's resolution and optimize deconvolution, measure a separate PSF for each laser wavelength, since resolution varies with this parameter. Finally, measure the PSF inside the FEP tube, as it may change due to refractive-index mismatch with the embedded gel. 

\textit{Generation 2 - Sample holders and mounting:} 
Model organisms such as zebrafish and Drosophila larvae can be mounted in the tube by standard syringe and needle aspiration (Figure \ref{fig:holders} (top)). Fixed samples are easiest to mount in a liquid like PBS, while live specimens must be mounted in an agarose gel (concentration $>$ 4$\%$) to arrest their motion. Mounting of small samples like gastruloids, organoids, and cell spheroids is more delicate and should be aspirated using a luer-lock system (Figure \ref{fig:holders} (bottom)). To minimize streaks in the images, clean and straighten tubes, as described in Section \ref{sec:generation2}B. 3D-printed holders are ideal for rapid prototyping but are prone to thermal drift, so for experiments with temperature control, we recommend using aluminum holders instead (see Section \ref{sec:generation5}C). Anodizing the aluminum holders and the temperature control unit (Section \ref{sec:generation4}C) ensures biocompatibility. We recommend black anodization as it minimizes light reflections and background glow in microscope images. However, a drawback of using an opaque viewing chamber is that it can be difficult to position the sample in the camera's field of view since it is hidden. One way to ease this task is to locate the tube first by using the strong scattering signal when the emission filter is removed, then translate along the tube with the stage to locate the sample.    

\textit{Generation 3 - Multi-channel imaging}:
To automate multi-channel imaging, we recommend integrating the stage, Arduino microcontroller, and camera in Micro-Manager. Interfacing the laser with the Arduino enables fast switching between the laser channels, preserving the fast imaging of SPIMs. We recommend mounting a Screwshield on the Arduino to provide additional ground (GND) connection points and to ensure more robust wiring. To ensure smooth operation without lag or unexpected shutdown, we suggest minimizing the file size of Micro-Manager configuration files. Finally, to ensure the laser sheet scans the entire specimen during automated multi-view imaging, we recommend z-stack imaging across the full width of the FEP tube.    

\textit{Generation 4 and 5 - Environmental control:}
To ensure a uniform temperature in the viewing chamber, we recommend building it in aluminum because of this material's excellent heat-transfer properties. We also recommend insulating the chamber's bottom surface with an acrylic block to prevent heat loss to the optical table. 

To ensure a steady flow rate, the flow path must be leak-free and bubble-free; for this, we recommend using Luer-lock syringes and fittings. A gas-tight system also maintains a constant $\text{CO}_2$ level, which controls the media's $ \text{pH} $ level. The pH level can be visualized using a pH-indicator dye or measured more accurately with a pH meter. Further leveraging aluminum's high heat-transfer coefficient, as in the viewing chamber, we also recommend using this material for the sample holder to ensure the flowing media reaches physiological temperature by the time it reaches the sample.

\begin{figure}[h]
    \centering
    \includegraphics[width=0.5\linewidth]{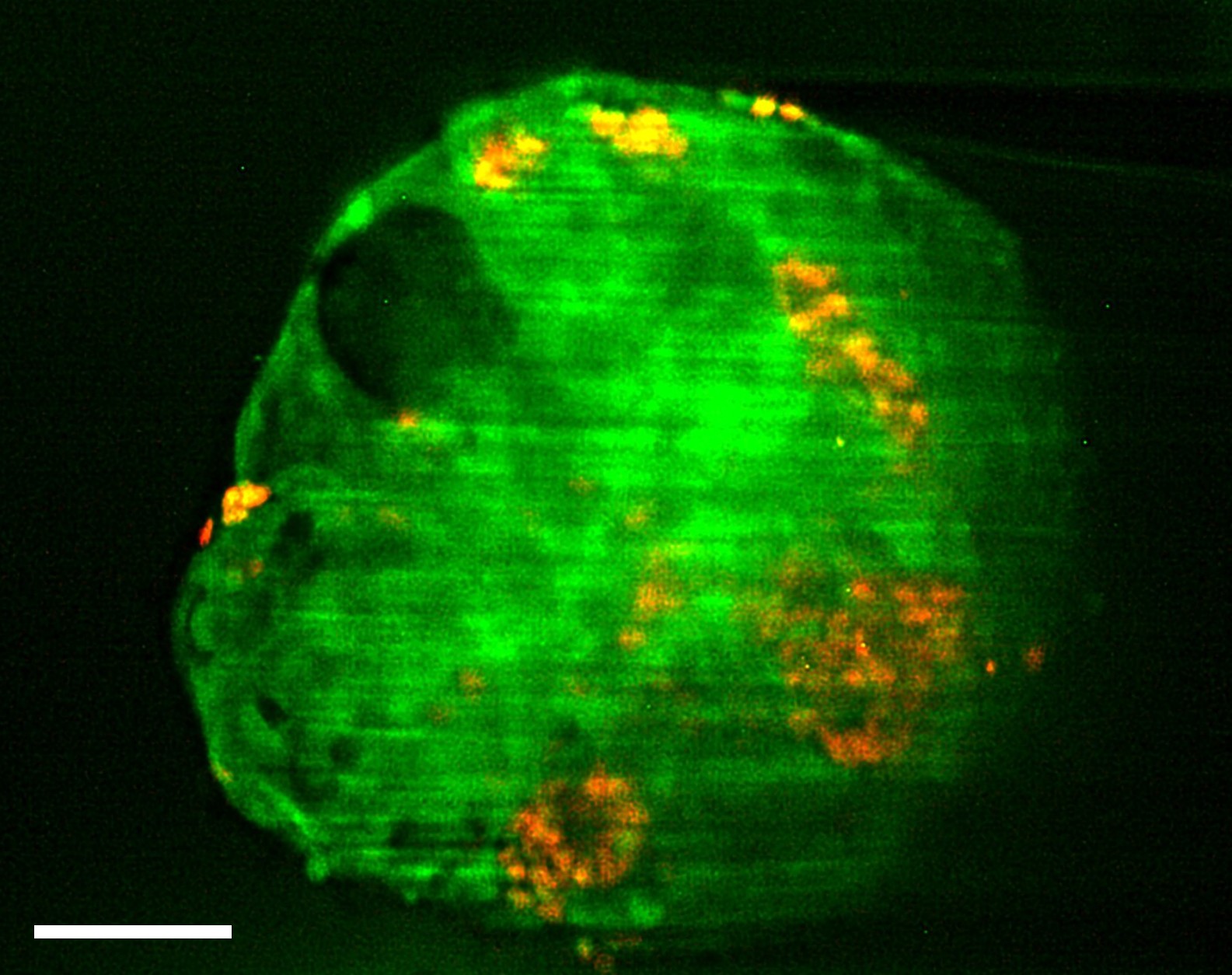}
    \caption{Optical section of a live gastruloid at 4 days post-aggregation (4 dpa), imaged 30 minutes after mounting in the capillary tube. The green signal corresponds to T/Bra GFP (mesoderm) reporters, and the red signal corresponds to Sox17 RFP (endoderm) reporters. The image has been deconvolved in Generic Deconvolution. The light sheet comes from the left. Scale bar: 100\,$\mu$m.}
    \label{fig:livegastruloid_decon}
\end{figure}

\textbf{Further improving the setup and contributions to the field:}
Close inspection of the stripes in Figure \ref{fig:livegastruloid_decon} shows that the light sheet points slightly downward, and this is because the beam hits the illumination objective slightly above its center. Replacing the 25 mm posts holding the mirrors with 20 mm posts would bring the beam closer to the center of the illumination objective and should remove the tilt, making the light sheet parallel to the focal plane. Without modifying the optical setup presented in this paper, the stripes in Figure \ref{fig:livegastruloid_decon} can be removed by vertically scanning the sample during image acquisition and then merging the images \cite{dong2014vertically}, or by imaging from multiple sides and by performing subsequent fusion \cite{swoger2007multi}. Stripes in SPIM images can also be removed in post-processing, as reviewed by \citet{ricci2022removing}. 

%

%The stripes in the image can be removed e.g. by pivoting the light sheet in multidirectional SPIM (mSPIM) \cite{huisken2007even} using a galvanic mirror,

%(Likewise, the quality of 3D images can be improved by increasing the number of acquisition angles prior to fusion since the best (typically, brightest) voxels from all acquisition angles are used in the fused 3D image.)

%The gastruloids used in this study emit autofluorescence when illuminated with the blue laser, which interferes with the Chi-induced Brachyury-GFP reporter signal and makes it difficult to localize the true GFP signal. To localize the GFP signal, the autofluorescence can be removed from the images, and can be quantified by considering the grayscale intensity in microscope images of control gastruloids not treated with Chi (which do not express Brachyury-GFP signal).

Biological specimen commonly emit autofluorescence when illuminated with a blue laser. For our gastruloids, such autofluorescence interferes with the Chi-induced Brachyury-GFP signal, making precise GFP localization difficult. To isolate the true GFP signal, the autofluorescence can be removed from the images, as quantified by imaging control gastruloids not treated with Chi (which do not express Brachyury-GFP).

%In our imaging experiments, we imaged samples saw few signs of light scattering from the samples, except for aggregates of cellulose nanocrystals, which we identified as bright spots in the microscope images. 

In our experiments, illumination with the blue (488 nm) laser produced reflections in the microscope images from the tube walls and the steel pin used to hold the samples, probably because the emission filter did not completely block the incident (blue) laser light. However, these reflections did not degrade the microscope images because they were generally far from the sample. Instead, because the reflections were recognizable features at different viewing angles, we used them to our advantage, as they aided the multi-view fusion step. This trick was particularly useful when fusing multi-view images of round 3 dpa gastruloids without distinct features, since these samples look very similar from different angles. 

In diffraction-limited optical systems like the FlowSPIM, the resolution along the optical (z) axis is lower than the resolution in the image (x,y) plane, as quantified by the PSF dimensions at FWHM (PSF at FWHM in x, y, z: 1.5 $\mu$m, 2.1 $\mu$m, 6.0 $\mu$m in our case). This anisotropy causes round objects to appear elongated in the z-direction. Fortunately, multi-view acquisition and fusion \cite{wu2016simultaneous,preibisch2014efficient}, as used in this paper, improves z-resolution so the true shape of 3D objects is recovered.

%In diffraction limited optics system such as the FlowSPIM, the extent of the PSF is longer along the optical axis (z) than in the image plane (x, y; PSF at FWHM in x, y, and z: 1.5 $\mu$m, 2.1 $\mu$m, 6.0 $\mu$m), resulting in poorer resolution in the z-direction and causing round objects to appear elongated in the z-direction. Luckily, multi-view acquisition and fusion \cite{wu2016simultaneous,preibisch2014efficient} as demonstrated in this paper makes the PSF more isotropic, thereby improving the z-resolution so that the true shape of 3D objects is recovered.  

To obtain a good measurement of the PSF, the beads should be smaller than the lateral resolution of the optical system, which is $\text{0.61}\lambda_{laser}/\text{NA}$=$\text{0.61}\cdot 0.488 \mu m/0.5$=$0.6 \mu m$ for the blue (488 nm) laser used in our system. Thus, using 0.5 $\mu$m beads to obtain the PSF, as in our experiments, satisfies the Rayleigh criterion given above. However, bead size also contributes to the measured PSF, and using smaller beads (e.g., 200 nm in mean diameter) would yield a better PSF, potentially improving the deconvolution result. The deconvolution quality could also be improved by measuring a separate PSF for each laser wavelength, since it varies with this parameter. To further improve the PSF measurement, measure it inside the sample tube to account for the slight refractive index mismatch with the agarose gel in which the beads are mounted. Finally, the PSF measurement can be improved by decreasing the microscope stage z-step size, since the PSF in this direction is resolved by only four optical sections ($PSF_{z,FWHM}=6.0 \mu m$, z-steps: $1.5 \mu$m). However, since we already operate at the stage's resolution limit, improving the z-resolution would require acquiring a new stage.  

%There are several ways to enhance the acquisition speed in live recordings, and they all come with their pros and cons (alone or in combination). One route is to increase the shutter speed; however the increased laser power required to obtain sufficient optical contrast can lead to lethal photodamage. Another strategy is to image from only one side instead of from multiple angles; however single side imaging often suffers from striping artifacts and light scattering. A third strategy and is to increase the z-step size  

%The paragraphs above describe how to improve spatial resolution. However, increasing the temporal resolution can be just as important, especially in live imaging experiments. 
%Below we suggest a few strategies, which can be combined. The first and simplest strategy is to increase the shutter speed. Here, one must make sure the increased laser power required to obtain sufficient optical contrast does not induce lethal photodamage. A second strategy is to image from only one side; however, expect striping artifacts and light scattering, especially when imaging deep into tissue. A third viable strategy is to increase the z-step size. To resolve features in the z-direction, one must meet the Nyquist sampling criterion, which here means the step size should not exceed half the PSF dimension in z at FWHM (6.0 $\mu$m in our case). As such, the step size should not exceed 3.0 $\mu$m in our setup. 

In SPIM systems, image acquisition can be accelerated by adding additional illumination and detection objectives to the viewing chamber to enable imaging or illumination from two sides simultaneously, thereby reducing the number of sample rotations in multi-view acquisition. For example, the mSPIM \cite{Huisken2007mSPIM} features two-sided illumination to reduce striping artifacts when imaging zebrafish embryos without rotating the sample. The MuViSPIM \cite{Krzic2012MuViSPIM} and the X-OpenSPIM \cite{Girstmair2022XOpenSPIM} add an additional detection objective to the mSPIM configuration, which drastically reduces the number of sample rotations when imaging Drosophila and flour beetles, respectively. However, unlike our setup, none of these setups had a temperature or flow control unit and are ill suited to imaging mammalian tissue cultures.  

%while implementing dual sided illumination and detection functionality to the FlowSPIM would increase the acquisition speed, adding the additional objectives would not be compatible with the heating chamber described in Section \ref{sec:generation4}. 

The emerging method of maintaining the viability of tissue cultures for long light sheet imaging experiments is to mount them in open chambers inside of an environmentally controlled incubation chamber, as utilized for organoid imaging \cite{deMedeiros2022MultiscaleOrganoid, Wysmolek2022MinimalComplexityLSFM, Beghin2022JeWells, Jain2025BrainOrganoidMorphodynamics}, and for gastruloid imaging \cite{Moos2024DualViewGastruloids,Suppinger2023Gastruloids}. This open-chamber incubation method has enabled researchers to image live samples over several hours and sometimes days; however, because the sample cannot be rotated, these setups do not support multi-view imaging. As such, imaging depth is usually limited to a few hundred microns before scattering becomes limiting, and it is best suited for small samples. In contrast, FlowSPIM offers both environmental control (temperature and flow) and multi-view imaging, combining the main advantages of previous SPIM and light sheet configurations.

%To obtain a z stack of the sample, most of these setups scan the sample through a horizontal light sheet and image from below   

\section{Conclusion}
\label{sec:conclusion}
This paper describes the iterative process of designing and building a light sheet microscope, and adapting it to different samples and experimental requirements. Starting from a stripped-down OpenSPIM microscope, we build upon the initial design by adding new hardware such as optical components, new laser lines, and a new flow-based incubation system. For each of the five generations of our setup, we describe the design requirements, the experimental challenges and pitfalls, and we present recommended experimental procedures. We hope this paper serves as a guide for researchers interested in building their own light sheet microscope, and that some of our technological innovations interest readers beyond the imaging community.   

\section*{Electronic material}

%Videos are available at \newline . 

Electronic material is available in Ref. \cite{Mossige_2026_zenodo_23898349} (\url{https://doi.org/10.5281/zenodo.23898349}): videos and full-resolution figures, CAD files of the machined and 3D-printed parts, the bill of materials and the table of changes between prototypes, the Arduino firmware, the Micro-Manager configuration files, raw and deconvolved images of biological specimens and cellulose nanocrystals, the point spread function, and the fluorescent-bead image stack used to measure the point spread function.

\section*{Author contributions}
\textbf{Endre Joachim Lerheim Mossige}: Conceptualization, Writing – original draft, Data curation, Formal analysis, Investigation, Methodology, Project administration, Resources, Software, Validation, Visualization, Writing – review \& editing. \textbf{Sergey Ponomartsev}: Conceptualization, Investigation, Methodology, Resources, Validation, Writing – review \& editing. \textbf{Natalia Smirnova}: Investigation, Methodology, Resources, Validation, Writing – review \& editing, Conceptualization. \textbf{Wietske van der Ent}: Conceptualization, Investigation, Methodology, Resources, Validation, Writing – review \& editing. \textbf{Kesavan Sekar}: Conceptualization, Investigation, Methodology, Resources, Validation, Formal analysis, Writing – review \& editing. \textbf{Siri Andresen}: Conceptualization, Investigation, Methodology, Resources, Validation, Writing – review \& editing. \textbf{Xian Hu}: Data curation, Formal analysis, Resources, Software, Validation, Visualization, Writing – review \& editing.
\textbf{Felix Margadant}: 
Resources, Writing – review \& editing, Methodology. \textbf{Rainer Heintzmann}: 
Resources, Software, Writing – review \& editing, Methodology. \textbf{Roland Kádár}: Conceptualization, Investigation, Methodology, Resources, Supervision, Writing – review \& editing. \textbf{Helene Knævelsrud}: Conceptualization, Supervision, Writing – review \& editing, Resources, Methodology. \textbf{Camila Vicencio Esguerra}: Conceptualization, Resources, Writing – review \& editing, Methodology. \textbf{Stefan Krauss}: Conceptualization, Funding acquisition, Project administration, Resources, Writing – review \& editing, Supervision. \textbf{Dag Kristian Dysthe}: Conceptualization, Funding acquisition, Investigation, Methodology, Project administration, Resources, Supervision, Writing – review \& editing. \textbf{Alexander Refsum Jensenius}: Conceptualization, Funding acquisition, Methodology, Project administration, Resources, Supervision, Writing – original draft, Writing – review \& editing.

\section*{Acknowledgments}
We thank Thomas Combriat and Ali Aslan Demir for help with computer interfacing and Micro-Manager implementation. We thank Jan Kristiansen, Hans Borg, Jonas Ringnes, and Bjørn Karsten Eriksen from the machine shop at the Department of Physics, University of Oslo, for machining sample holders, viewing chambers, and steel pins to seal the tubes, and William Nævdal Sigurdsson and Ole Dorholt from the electronics lab at Dept. Physics, Univ. Oslo for hardware interfacing between the Arduino microcontroller and the laser. We thank Vishesh Kumar Dubey and Yi Hu for help with laser alignment, and Balpreet Singh Ahluwalia and Kay Oliver Shrink for useful inputs on laser alignment and optics. We also thank the NorMIC Imaging Platform at the Department of Biosciences, University of Oslo, for providing assistance and access to the confocal microscope for quality control of gastruloids, and Frode Miltzow Skjeldal for training EJLM and SP.  

We thank Gabor Juhasz, HUN-REN Szeged, and Thomas P. Neufeld, University of Minnesota, for generously sharing reagents, and Iftach Nachman, Tel-Aviv University, Israel, for providing the mouse embryonic stem cell line.

This project is part of the UiO: Life Science convergence environment Integrated Technologies for Tracking Organoid Morphogenesis (ITOM), with contributions from the convergence environment AUTORHYTHM. Convergence environments are interdisciplinary research groups that aim to solve grand challenges related to health and the environment. UiO’s interdisciplinary strategic area UiO:Life Science funds them. 

%The project also received funding from three Norwegian centres of excellence, namely PoreLab, Hybrid Technology Hub, and RITMO Centre for Interdisciplinary Studies in Rhythm, Time and Motion. 

This work was partly supported by the Research Council of Norway through its Centres of Excellence funding scheme, project number 262652. Views and opinions expressed are, however, those of the author(s) only and do not necessarily reflect those of the European Union or the European Research Council Executive Agency. Neither the European Union nor the granting authority can be held responsible.

EJLM acknowledges support from UiO Growth House, Kristine Bonnevie's reisestipend, and from the Bridging Nordic Microscopy Infrastructure Short-Term Scientific Missions program. SK acknowledges support from the European Innovation Council (Pathfinder Grant, Supervised Morphogenesis in Gastruloids, Grant No. 101071203). H.K. was supported by funding from the European Union (ERC, FINALphagy, 101039174). 

\section*{Appendix}

\subsection*{Mouse Embryonic Stem Cell Culture and Generation of Gastruloids}
In this research, we used the E14 Bra-GFP:Sox17-RFP mouse embryonic stem cell (mESC) line~\citep{pour2022emergence}, kindly provided by Iftach Nachman (Tel-Aviv University, Israel).

\subsubsection*{Cell Culture}
mESC were grown on wells treated with gelatin (ES-006-B, Merck) in serum + LIF medium. The medium consisted of knockout DMEM (10829018, Gibco) supplemented with 15\% fetal bovine serum (10309433, HyClone), 1x non-essential amino acids (11140050, Gibco), 1x Gluta-MAX (35050061, Gibco), 100 $\mu$m beta-mercaptoethanol (M3148, Merck), and 1:10000 mouse leukemia inhibitory factor (LIF) (ESG1107, Merck). We grew mESCs for 2--3 passages before gastruloid aggregation. We passaged cells every second day with Accutase (A6964, Merck).

\subsubsection*{Gastruloid Generation}
To generate gastruloids, we used a previously published protocol \cite{BaillieJohnson2015}. Briefly, 300 mESC were plated in 40 µL home-made N2B27 medium (DMEM/F12 (11039021, Gibco) and Neurobasal (12348017, Gibco) in a 1:1 ratio, supplemented with 0.5x N2 (17502048, Gibco), 0.5x B27 (17504044, Gibco), 0.5x GlutaMAX (35050061, Gibco), 1x non-essential amino acids (11140050, Gibco), 1x Na-Pyruvate (11360039, Gibco), 50 $\mu$m beta-mercaptoethanol (M3148, Merck), and 0005\% BSA (A8412, Merck)) into a 96-well U-bottom cell-repellent microplate (650970, Greiner Bio-One). After 48 h of aggregation, add 150 µL of N2B27 supplemented with 3 µg/mL CHIR 99021 (4423, Tocris) to the wells. Then, every 24 h, remove 150 µL of media and replace it with fresh N2B27. Cells and gastruloids were maintained at 37°C, 5\% CO2, and 100\% humidity.

\subsubsection*{Fixation}
Gastruloids were fixed in 4\% paraformaldehyde (PFA) (158127, Merck) diluted in DPBS (14190094, Gibco) for 1 h at room temperature on a shaking platform. They were then washed three times in DPBS and stored at 4°C in DPBS.

\subsection*{Zebrafish sample preparation}

\subsubsection*{Zebrafish husbandry}
The transgenic zebrafish line Tg(Fli:GFP) was maintained according to standard protocols and local animal welfare regulations. We obtained eggs by natural spawning and maintained them at 28°C in an incubator with a 14/10h light-dark cycle until further use. The Norwegian Food Safety Authority (FOTS \#27539) approved all zebrafish experiments.

\subsubsection*{Cell culture}
U-251 cells were maintained at 37°C with 5\% CO$_2$ in DMEM (11995065, Gibco) containing 10\% fetal bovine serum (A5256701, Gibco). Cells were passaged twice a week by rinsing attached cells with DPBS (14190094, Gibco), then detaching them with 0.25\% Trypsin-EDTA (25200056, Gibco). Cells were grown to \textasciitilde70\% confluence before engraftment.

\subsubsection*{Engraftments}
We performed engraftments as described previously \cite{ding2025automated}. In brief, we labeled detached cells with CM-DiI (C7000, Invitrogen) according to the manufacturer’s instructions and suspended them in DPBS (14190094, Gibco) to a concentration of $1 \times 10^7$ cells/mL. We anesthetized dechorionated zebrafish embryos (2 dpf) with 40\,µg/mL MS-222 (E10521, Sigma-Aldrich) and placed them on 2\% agarose (A9539, Sigma-Aldrich) in E3 medium. We engrafted approximately 100--300 cells into the hindbrain ventricle of each embryo. After engraftment, we transferred embryos to fresh E3, incubated them for 1\,h at 28°C, then placed them at 34°C in an incubator with a 14/10h light-dark cycle until further procedures.

\subsubsection*{Fixation}
Larvae were anesthetized with 40\,µg/mL MS-222 (E10521, Sigma-Aldrich) and fixed with 4\% PFA (158127, Merck) diluted in DPBS (14190094, Gibco), overnight at 4°C on a shaking platform. Samples were washed three times in DPBS before storage at 4°C in DPBS.

\subsection*{Drosophila sample preparation}

\subsubsection*{Fly strains and maintenance}

\textit{Drosophila} lines used in this study were \textit{UAS-Atg7 RNAi TRiP.JF02787} (Bloomington Drosophila Stock Center, 27707), \textit{cg-Gal4, FRT42D, UAS-myrRFP; UAS-GFP-Atg8a} \cite{Juhasz2008} kindly provided by Thomas P. Neufeld, and \textit{hsflp; 3xmCherry-Atg8, UAS-GFP/CyO; Act$>$CD2$>$Gal4, UAS-Dicer2/TM6} \cite{Hegedus2016} kindly provided by Gábor Juhász.

Fly strains were kept and raised in bottles containing standard potato mash fly food: 27.3\,g/liter dry yeast (LeSaffre 41,036), 32.7\,g/liter dried potato powder (Hoff 24,458), 60\,g/liter sucrose (NordicSugar AS 10,330), 0.73\% agar (AS PALS 77,000), 0.2\% methyl 4-hydroxybenzoate (Sigma-Aldrich, H5501) dissolved in ethanol, and 0.45\% propionic acid (Sigma-Aldrich, P1386).

\subsubsection*{Sample preparation}

The desired fly lines were crossed and kept at 25$^\circ$C, with egg deposition allowed for a period of 4\,h before the adult flies were removed and the eggs were incubated for a total of 96\,h. Afterward, third-instar (L3) feeding larvae were washed, placed in 20\% sucrose (Sigma-Aldrich, S7903) diluted in PBS, and starved for 4 hours. Before imaging, the larvae were anesthetized by exposing them to ether (Merck, 1.00921.1000) for 3 minutes in a glass beaker.

\subsection*{Cellulose nanocrystal suspension preparation}

Cellulose nanocrystals (CelluForce NCC\textsuperscript{\textregistered} NCV100-NASD90) were dispersed in Milli-Q water at 4 wt\% and sonicated for 30 min at 60\% amplitude and a 60\% duty cycle using a Branson Sonifier 450 equipped with a 20 kHz ultrasonic horn and a 1/2-inch-diameter probe. The sample was maintained in an ice bath, which was replenished regularly to minimize heating. After two days, the suspension separated into a clear isotropic upper phase and a turbid lower phase enriched in liquid-crystalline domains with chiral nematic ordering. We collected the turbid phase and used it for subsequent experiments.

\subsection*{Timescales of heat transfer, gas diffusion and flow}

This section presents estimates of typical transport timescales associated with flow-enhanced incubation of 3D cell cultures. We use this to optimize the flow rate of media over the sample. 

\subsubsection*{Heat transfer through the FEP-tube}
The characteristic time scale of heat diffusion through the capillary (FEP) tube, from the water bath to the flowing media, depends on the thermal diffusivity of the FEP tube, $\alpha_T$, and the half-width of the FEP tube, $w_{1/2}$. In our setup, $w_{1/2}=0.8\,mm$ and $\alpha_T=0.107\, mm^2/s$ \cite{olifirov2021tribological}, which gives

\begin{equation}
    \tau_T \sim \frac{w_{1/2}^2}{\alpha_T} = 6\, s.
\end{equation} 

\subsubsection*{$CO_2$ diffusion from the cell aggregate to the media}
The characteristic time scale of $CO_2$ diffusion depends on the diffusivity of $CO_2$ in water (cell media contains mostly water), $\alpha_{CO_2}$, and the size of the cell aggregate, $a$. In our setup, $\alpha_{CO_2}=2.9020 \cdot 10^{-3}\,mm^2/s$\cite{cadogan2014diffusion} and $a\approx 0.4\,mm$, which gives

\begin{equation}
    \tau_{CO_2} \sim \frac{a^2}{\alpha_{CO_2}} = 55\, s,
\end{equation} 

so one order of magnitude slower than the timescale of heat diffusion. 

\subsubsection*{Timescale of flow past the sample}
To maintain cell viability, we must ensure the flow efficiently advects $ CO_2$ away from the sample. This means the flow timescale past the spheroid, $\tau_u$, must be faster than the gas diffusion timescale, i.e. $\tau_u \ll \tau_{C0_2}$. The minimal velocity (giving $\tau_u = \tau_{C0_2}$, and so $Pe=1$) is simply

\begin{equation}
    u = \frac{a}{\tau_{CO_2}} = \frac{\alpha_{CO_2}}{a} = 0.0073\, mm/s = 7.3\, \mu m /s.
\end{equation} 

These calculations are order-of-magnitude estimates.

% Appendix: Bill of Materials for the Final FlowSPIM Prototype
% Non-tabular LaTeX-compatible format.

\nocite{*}

\bibliography{apssamp}% Produces the bibliography via BibTeX.

@PREAMBLE{
 "\providecommand{\noopsort}[1]{}" 
 # "\providecommand{\singleletter}[1]{#1}%" 
}

@incollection{weber2014light,
  title={Light sheet microscopy},
  author={Weber, Michael and Mickoleit, Michaela and Huisken, Jan},
  booktitle={Methods in cell biology},
  volume={123},
  pages={193--215},
  year={2014},
  doi={10.1016/B978-0-12-420138-5.00011-2},
  publisher={Elsevier}
}

@article{girkin2018light,
  title={The light-sheet microscopy revolution},
  author={Girkin, John M and Carvalho, Mariana Torres},
  journal={Journal of Optics},
  volume={20},
  number={5},
  pages={053002},
  year={2018},
  publisher={IOP Publishing}
}

@article{stelzer2021light,
  title={Light sheet fluorescence microscopy},
  author={Stelzer, Ernst HK and Strobl, Frederic and Chang, Bo-Jui and Preusser, Friedrich and Preibisch, Stephan and McDole, Katie and Fiolka, Reto},
  journal={Nature Reviews Methods Primers},
  volume={1},
  number={1},
  pages={73},
  year={2021},
  doi={10.1038/s43586-021-00069-4},
  publisher={Nature Publishing Group UK London}
}

@article{Keller2008ZebrafishDSLM,
  author    = {Philipp J. Keller and Annette D. Schmidt and Joachim Wittbrodt and Ernst H.K. Stelzer},
  title     = {Reconstruction of zebrafish early embryonic development by scanned light sheet microscopy},
  journal   = {Science},
  year      = {2008},
  volume    = {322},
  number    = {5904},
  pages     = {1065--1069},
  doi       = {10.1126/science.1162493}
}

@article{Huisken2009ReviewSPIM,
  author    = {Jan Huisken and Didier Y.R. Stainier},
  title     = {Selective plane illumination microscopy techniques in developmental biology},
  journal   = {Development},
  year      = {2009},
  volume    = {136},
  number    = {12},
  pages     = {1963--1975},
  doi       = {10.1242/dev.022426}
}

@article{Ahrens2013ZebrafishBrain,
  author    = {Misha B. Ahrens and Michael B. Orger and Drew N. Robson and Jennifer M. Li and Philipp J. Keller},
  title     = {Whole-brain functional imaging at cellular resolution using light-sheet microscopy},
  journal   = {Nature Methods},
  year      = {2013},
  volume    = {10},
  number    = {5},
  pages     = {413--420},
  doi       = {10.1038/nmeth.2434}
}

@article{Preibisch2008DrosophilaSPIM,
  author    = {Stephan Preibisch and Robert Ejsmont and Tobias Rohlfing and Pavel Tomancak},
  title     = {Towards digital representation of {D}rosophila embryogenesis},
  journal   = {Proceedings of ISBI},
  year      = {2008},
  pages     = {324--327},
  doi       = {10.1109/ISBI.2008.4541033}
}

@article{Krzic2012Multiview,
  author    = {Uros Krzic and Steffen G{\"u}nther and Thomas E. Saunders and Stephan J. Streichan and Lars Hufnagel},
  title     = {Multiview light-sheet microscope for rapid in toto imaging},
  journal   = {Nature Methods},
  year      = {2012},
  volume    = {9},
  number    = {7},
  pages     = {730--733},
  doi       = {10.1038/nmeth.2064}
}

@article{Schmied2016DrosophilaProtocol,
  author    = {Christopher Schmied and Pavel Tomancak},
  title     = {Sample preparation and mounting of {D}rosophila embryos for multiview light sheet microscopy},
  journal   = {Methods in Molecular Biology},
  year      = {2016},
  volume    = {1478},
  pages     = {189--202},
  doi       = {10.1007/978-1-4939-6371-3_10}
}

@article{Huisken2007mSPIM,
  author  = {Jan Huisken and Didier Y. R. Stainier},
  title   = {Even fluorescence excitation by multidirectional selective plane illumination microscopy (mSPIM)},
  journal = {Optics Letters},
  year    = {2007},
  volume  = {32},
  number  = {17},
  pages   = {2608--2610},
  doi     = {10.1364/OL.32.002608}
}

@article{kaufmann2012multilayer,
  title={Multilayer mounting enables long-term imaging of zebrafish development in a light sheet microscope},
  author={Kaufmann, Anna and Mickoleit, Michaela and Weber, Michael and Huisken, Jan},
  journal={Development},
  volume={139},
  number={17},
  pages={3242--3247},
  year={2012},
  doi={10.1242/dev.082586},
  publisher={Company of Biologists}
}

@article{Chen2024PropsLSFM,
  author    = {Bingying Chen and Bo-Jui Chang and Stephan Daetwyler and others},
  title     = {Projective light-sheet microscopy with flexible parameter selection},
  journal   = {Nature Communications},
  year      = {2024},
  volume    = {15},
  pages     = {2755},
  doi       = {10.1038/s41467-024-46693-y}
}

@article{van2014symmetry,
  title={Symmetry breaking, germ layer specification and axial organisation in aggregates of mouse embryonic stem cells},
  author={Van den Brink, Susanne C and Baillie-Johnson, Peter and Balayo, Tina and Hadjantonakis, Anna-Katerina and Nowotschin, Sonja and Turner, David A and Martinez Arias, Alfonso},
  journal={Development},
  volume={141},
  number={22},
  pages={4231--4242},
  year={2014},
  doi={10.1242/dev.113001},
  publisher={The Company of Biologists}
}

@article{ho2025spontaneous,
  title={Spontaneous elongation of 3{D} gastruloids from local cell polarity alignment},
  author={Ho, Richard DJG and Mossige, Endre JL and Ponomartcev, Sergei and Smirnova, Natalia and Hu, Xian and Dai, Keqing Sunny and Krauss, Stefan and Dysthe, Dag Kristian and Angheluta, Luiza},
  journal={arXiv preprint arXiv:2509.08929},
  url={
https://doi.org/10.48550/arXiv.2509.08929
},
  year={2025}
}

@article{van20213d,
  title={3{D} gastruloids: a novel frontier in stem cell-based in vitro modeling of mammalian gastrulation},
  author={van den Brink, Susanne C and van Oudenaarden, Alexander},
  journal={Trends in cell biology},
  volume={31},
  number={9},
  pages={747--759},
  year={2021},
  url={https://www.cell.com/trends/cell-biology/fulltext/S0962-8924(21)00123-9},
  publisher={Elsevier}
}

@article{weiss2021tutorial,
  title={Tutorial: practical considerations for tissue clearing and imaging},
  author={Weiss, Kurt R and Voigt, Fabian F and Shepherd, Douglas P and Huisken, Jan},
  journal={Nature protocols},
  volume={16},
  number={6},
  pages={2732--2748},
  year={2021},
  doi={10.1038/s41596-021-00502-8},
  publisher={Nature Publishing Group UK London}
}

@article{kaltenecker2024deep,
  title={Deep learning and {3D} imaging reveal whole-body alterations in obesity},
  author={Kaltenecker, Doris and Horvath, Izabela and Al-Maskari, Rami and Kolabas, Zeynep Ilgin and Chen, Ying and Hoeher, Luciano and Todorov, Mihail and Kapoor, Saketh and Ali, Mayar and Kofler, Florian and others},
  journal={bioRxiv},
  pages={2024--08},
  year={2024},
  doi={10.1101/2024.08.18.608300},
  publisher={Cold Spring Harbor Laboratory}
}

@article{BaillieJohnson2015,
  title={Generation of aggregates of mouse embryonic stem cells that show symmetry breaking, polarization and emergent collective behaviour in vitro},
  author={Baillie-Johnson, Peter and Van den Brink, Susanne Carina and Balayo, Tina and Turner, David Andrew and Arias, Alfonso Martinez},
  journal={Journal of visualized experiments: JoVE},
  number={105},
  pages={53252},
  doi={10.3791/53252},
  year={2015}
}

@article{pitrone2013openspim,
  title={Open{SPIM}: an open-access light-sheet microscopy platform},
  author={Pitrone, Peter G and Schindelin, Johannes and Stuyvenberg, Luke and Preibisch, Stephan and Weber, Michael and Eliceiri, Kevin W and Huisken, Jan and Tomancak, Pavel},
  journal={Nature methods},
  volume={10},
  number={7},
  pages={598--599},
  year={2013},
  doi={10.1038/nmeth.2507},
  publisher={Nature Publishing Group US New York}
}

@article{girstmair2022time,
  title={Time to upgrade: a new {OpenSPIM} guide to build and operate advanced {OpenSPIM} configurations},
  author={Girstmair, Johannes and Moon, HongKee and Brillard, Charl{\`e}ne and Haase, Robert and Tomancak, Pavel},
  journal={Advanced biology},
  volume={6},
  number={4},
  pages={2101182},
  year={2022},
  doi={10.1002/adbi.202101182},
  publisher={Wiley Online Library}
}

@article{girstmair2016light,
  title={Light-sheet microscopy for everyone? Experience of building an Open{SPIM} to study flatworm development},
  author={Girstmair, Johannes and Zakrzewski, Anne and Lapraz, Fran{\c{c}}ois and Handberg-Thorsager, Mette and Tomancak, Pavel and Pitrone, Peter Gabriel and Simpson, Fraser and Telford, Maximilian J},
  journal={BMC Developmental Biology},
  volume={16},
  number={1},
  pages={22},
  year={2016},
  publisher={Springer}
}

@article{voigt2019mesospim,
  title={The meso{SPIM} initiative: open-source light-sheet microscopes for imaging cleared tissue},
  author={Voigt, Fabian F and Kirschenbaum, Daniel and Platonova, Evgenia and Pag{\`e}s, St{\'e}phane and Campbell, Robert AA and Kastli, Rahel and Schaettin, Martina and Egolf, Ladan and Van Der Bourg, Alexander and Bethge, Philipp and others},
  journal={Nature methods},
  volume={16},
  number={11},
  pages={1105--1108},
  year={2019},
  doi={10.1038/s41592-019-0554-},
  publisher={Nature Publishing Group US New York}
}

@article{edelstein2010computer,
  title={Computer control of microscopes using $\mu${M}anager},
  author={Edelstein, Arthur and Amodaj, Nenad and Hoover, Karl and Vale, Ron and Stuurman, Nico},
  journal={Current protocols in molecular biology},
  volume={92},
  number={1},
  pages={14--20},
  year={2010},
  doi={10.1002/0471142727.mb1420s92},
  publisher={Wiley Online Library}
}

@misc{webpagekeyOpenSPIM,
  author = {OpenSPIM},
  title = {Laser alignment},
  howpublished = {\url{https://openspim.org/Alignment_of_laser}},
  year = {2025},
  note = {Accessed: 2026-02-16}
}

@misc{webpagekeyOpenSPIM_parts,
  author = {OpenSPIM},
  title = {Table of parts},
  howpublished = {\url{https://openspim.org/Table_of_parts}},
  year = {2025},
  note = {Accessed: 2026-08-20}
}

@misc{samplemounting,
  author = {Huisken lab},
  title = {Sample mounting},
  howpublished = {\url{https://huiskenlab.com/sample-mounting/}},
  year = {2026},
  note = {Accessed: 2026-04-30}
}

@article{huisken2004optical,
  title={Optical sectioning deep inside live embryos by selective plane illumination microscopy},
  author={Huisken, Jan and Swoger, Jim and Del Bene, Filippo and Wittbrodt, Joachim and Stelzer, Ernst HK},
  journal={Science},
  volume={305},
  number={5686},
  pages={1007--1009},
  year={2004},
  publisher={American Association for the Advancement of Science}
}

@article{pampaloni2014tissue,
  title={Tissue-culture light sheet fluorescence microscopy ({TC-LSFM}) allows long-term imaging of three-dimensional cell cultures under controlled conditions},
  author={Pampaloni, Francesco and Berge, Ulrich and Marmaras, Anastasios and Horvath, Peter and Kroschewski, Ruth and Stelzer, Ernst HK},
  journal={Integrative Biology},
  volume={6},
  number={10},
  pages={988--998},
  year={2014},
  publisher={Oxford University Press}
}

@article{lorenzo2011live,
  title={Live cell division dynamics monitoring in {3D} large spheroid tumor models using light sheet microscopy},
  author={Lorenzo, Corinne and Frongia, C{\'e}line and Jorand, Rapha{\"e}l and Fehrenbach, J{\'e}r{\^o}me and Weiss, Pierre and Maandhui, Amina and Gay, Guillaume and Ducommun, Bernard and Lobjois, Val{\'e}rie},
  journal={Cell division},
  volume={6},
  number={1},
  pages={22},
  year={2011},
  publisher={Springer}
}

@article{huisken2009selective,
  title={Selective plane illumination microscopy techniques in developmental biology},
  author={Huisken, Jan and Stainier, Didier YR},
  journal={Development},
  volume={136},
  number={12},
  year={2009},
  doi={10.1242/dev.022426},
  publisher={Oxford University Press for The Company of Biologists Limited}
}

@article{kriesi2019integrated,
  title={Integrated flow chamber system for live cell microscopy},
  author={Kriesi, Carlo and Steinert, Martin and Marmaras, Anastasios and Danzer, Claudia and Meskenaite, Virginia and Kurtcuoglu, Vartan},
  journal={Frontiers in bioengineering and biotechnology},
  volume={7},
  pages={91},
  year={2019},
  doi={10.3389/fbioe.2019.00091},
  publisher={Frontiers Media SA}
}

@misc{engineeringtoolbox_thermal_conductivity_metals,
  title        = {Thermal Conductivity of Metals},
  author       = {{Engineering Toolbox}},
  year         = {2026},
  howpublished = {\url{https://www.engineeringtoolbox.com/thermal-conductivity-metals-d_858.html}},
  note         = {Accessed: 2026-05-19}
}

@misc{engineeringtoolbox_thermal_conductivity,
  title        = {Thermal Conductivity},
  author       = {{Engineering Toolbox}},
  year         = {2026},
  howpublished = {\url{https://www.engineeringtoolbox.com/thermal-conductivity-d_429.html}},
  note         = {Accessed: 2026-05-19}
}

@article{huh20113d,
  title={From {3D} cell culture to organs-on-chips},
  author={Huh, Dongeun and Hamilton, Geraldine A and Ingber, Donald E},
  journal={Trends in cell biology},
  volume={21},
  number={12},
  pages={745--754},
  year={2011},
  url={https://www.cell.com/trends/cell-biology/fulltext/S0962-8924(11)00195-4},
  publisher={Elsevier}
}

@article{huh2010reconstituting,
  author  = {Huh, Dongeun and Matthews, Brigham D. and Mammoto, Akiko and Montoya-Zavala, Mario and Hsin, Hsiang-Yi and Ingber, Donald E.},
  title   = {Reconstituting Organ-Level Lung Functions on a Chip},
  journal = {Science},
  year    = {2010},
  volume  = {328},
  number  = {5986},
  pages   = {1662--1668},
  doi     = {10.1126/science.1188302}
}

@article{kim2012humanGutChip,
  author  = {Kim, Hyun Jung and Huh, Dongeun and Hamilton, Gerald and Ingber, Donald E.},
  title   = {Human Gut-on-a-Chip for Modeling Intestinal Inflammation and Drug Transport},
  journal = {Lab on a Chip},
  year    = {2012},
  volume  = {12},
  number  = {12},
  pages   = {2165--2174},
  doi     = {10.1039/c2lc40074j}
}

@article{zhang2016heartOnChip,
  author  = {Zhang, Y. S. and Arneri, A. and Bersini, S. and Shin, S.-R. and Zhu, K. and Goli-Malekabadi, Z. and Aleman, J. and Colosi, C. and Busignani, F. and Dell'Erba, V. and others},
  title   = {Bioprinting {3D} Microfibrous Scaffolds for Engineering Endothelialized Myocardium and Heart-on-a-Chip Devices},
  journal = {Biomaterials},
  year    = {2016},
  volume  = {110},
  pages   = {45--59},
  doi     = {10.1016/j.biomaterials.2016.09.003}
}

@article{bauwens2014liverOnChip,
  author  = {Bhatia, Sangeeta N. and Ingber, Donald E.},
  title   = {Microfluidic Organs-on-Chips},
  journal = {Nature Biotechnology},
  year    = {2014},
  volume  = {32},
  number  = {8},
  pages   = {760--772},
  doi     = {10.1038/nbt.2989}
}

@article{eschenhagen2017engineeredHeart,
  author  = {Eschenhagen, Thomas and Fink, Christian and Remmers, Udo and Scholz, Harald and Wattchow, James and Cassard, Alexandre and Fauvernier, Julien and others},
  title   = {Three-Dimensional Reconstitution of Embryonic Cardiomyocytes in a Perfused Microbioreactor for Cardiotoxicity Testing},
  journal = {Journal of Molecular and Cellular Cardiology},
  year    = {2017},
  volume  = {103},
  pages   = {12--22},
  doi     = {10.1016/j.yjmcc.2017.01.005}
}

@article{johns2022autofluorescence,
  title={Autofluorescence spectroscopy for quantitative analysis of cellulose nanocrystals},
  author={Johns, Marcus A and Abu-Namous, Jude and Zhao, Hongying and Gattrell, Michael and Lockhart, James and Cranston, Emily D},
  journal={Nanoscale},
  volume={14},
  number={45},
  pages={16883--16892},
  year={2022},
  doi={10.1039/D2NR04823J},
  publisher={Royal Society of Chemistry}
}

@article{kalita2015isolation,
  title={Isolation and characterization of crystalline, autofluorescent, cellulose nanocrystals from saw dust wastes},
  author={Kalita, E and Nath, BK and Agan, F and More, V and Deb, P},
  journal={Industrial Crops and Products},
  volume={65},
  pages={550--555},
  year={2015},
  doi={10.1016/j.indcrop.2014.10.004},
  publisher={Elsevier}
}

@article{gray2015chiral,
  title={Chiral nematic structure of cellulose nanocrystal suspensions and films; polarized light and atomic force microscopy},
  author={Gray, Derek G and Mu, Xiaoyue},
  journal={Materials},
  volume={8},
  number={11},
  pages={7873--7888},
  year={2015},
  doi={10.3390/ma8115427},
  publisher={MDPI}
}

@book{dunmur2010soap,
  title={Soap, science and flat-screen TVs: a history of liquid crystals},
  author={Dunmur, David and Sluckin, Tim},
  year={2010},
  publisher={Oxford University Press}
}

@misc{micromanagerCoboltOfficial,
  title        = {CoboltOfficial},
  author       = {{Cobolt Official}},
  year         = {2026},
  howpublished = {\url{https://micro-manager.org/CoboltOfficial}},
  note         = {Accessed: 2026-05-19}
}

@misc{micromanagerArduino,
  title        = {Arduino},
  author       = {{Nico Stuurman}},
  year         = {2026},
  howpublished = {\url{https://micro-manager.org/Arduino}},
  note         = {Accessed: 2026-05-19}
}

@article{pour2022emergence,
  title={Emergence and patterning dynamics of mouse-definitive endoderm},
  author={Pour, Maayan and Kumar, Abhishek Sampath and Farag, Naama and Bolondi, Adriano and Kretzmer, Helene and Walther, Maria and Wittler, Lars and Meissner, Alexander and Nachman, Iftach},
  journal={Iscience},
  volume={25},
  number={1},
  year={2022},
  doi={10.1016/j.isci.2021.103556},
  publisher={Elsevier}
}

@article{ding2025automated,
  title={Automated microinjection for zebrafish xenograft models},
  author={Ding, Yi and van der Kolk, Kees-Jan and van der Ent, Wietske and Scotto di Mase, Michele and Kowald, Saskia and Huizing, Jenny and Vidal Teuton, Jana M and Mishra, Gunja and Kempers, Maxime and Almter, Rusul and others},
  journal={npj Biomedical Innovations},
  volume={2},
  number={1},
  pages={13},
  year={2025},
  doi={10.1038/s44385-025-00016-y},
  publisher={Nature Publishing Group UK London}
}

@article{Juhasz2008,
  author = {Juh{\'a}sz, G{\'a}bor and Hill, Julia H. and Yan, Yue and Sass, Mikl{\'o}s and Baehrecke, Eric H. and Backer, John M. and Neufeld, Thomas P.},
  title = {The class {III} {PI(3)K V}ps34 promotes autophagy and endocytosis but not {TOR} signaling in {D}rosophila},
  journal = {The Journal of Cell Biology},
  year = {2008},
  volume = {181},
  number = {4},
  pages = {655--666},
  doi = {10.1083/jcb.200712051}
}

@article{Hegedus2016,
  author = {Heged{\H{u}}s, Katalin and Tak{\'a}ts, Szabolcs and Kov{\'a}cs, Attila L. and Juh{\'a}sz, G{\'a}bor},
  title = {Evolutionarily conserved role and physiological relevance of a {STX17}/{S}yntaxin17-containing {SNARE} complex in autophagosome fusion with endosomes and lysosomes},
  journal = {Molecular Biology of the Cell},
  year = {2016},
  volume = {27},
  number = {25},
  pages = {3268--3282},
  doi = {10.1091/mbc.E16-03-0205}
}

@article{olifirov2021tribological,
  title={Tribological, mechanical and thermal properties of fluorinated ethylene propylene filled with Al-Cu-Cr quasicrystals, polytetrafluoroethylene, synthetic graphite and carbon black},
  author={Olifirov, Leonid K and Stepashkin, Andrey A and Sherif, Galal and Tcherdyntsev, Victor V},
  journal={Polymers},
  volume={13},
  number={5},
  pages={781},
  year={2021},
  publisher={MDPI},
  doi={10.3390/polym13050781}
}

@article{cadogan2014diffusion,
  title={Diffusion coefficients of {CO}2 and {N}2 in water at temperatures between 298.15 {K} and 423.15 {K} at pressures up to 45 {MP}a},
  author={Cadogan, Shane P and Maitland, Geoffrey C and Trusler, JP Martin},
  journal={Journal of Chemical \& Engineering Data},
  volume={59},
  number={2},
  pages={519--525},
  year={2014},
  publisher={ACS Publications},
  doi={10.1021/je401008s}  
}

@article{huisken2007even,
  title={Even fluorescence excitation by multidirectional selective plane illumination microscopy (m{SPIM})},
  author={Huisken, Jan and Stainier, Didier YR},
  journal={Optics letters},
  volume={32},
  number={17},
  pages={2608--2610},
  year={2007},
  doi={10.1364/OL.32.002608},
  publisher={Optical Society of America}
}

@article{dong2014vertically,
  title={Vertically scanned laser sheet microscopy},
  author={Dong, Di and Arranz, Alicia and Zhu, Shouping and Yang, Yujie and Shi, Liangliang and Wang, Jun and Shen, Chen and Tian, Jie and Ripoll, Jorge},
  journal={Journal of biomedical optics},
  volume={19},
  number={10},
  pages={106001--106001},
  year={2014},
  doi={10.1117/1.JBO.19.10.106001},
  publisher={Society of Photo-Optical Instrumentation Engineers}
}

@article{swoger2007multi,
  title={Multi-view image fusion improves resolution in three-dimensional microscopy},
  author={Swoger, Jim and Verveer, Peter and Greger, Klaus and Huisken, Jan and Stelzer, Ernst HK},
  journal={Optics express},
  volume={15},
  number={13},
  pages={8029--8042},
  year={2007},
  doi={10.1364/OE.15.008029},
  publisher={Optical Society of America}
}

@article{ricci2022removing,
  title={Removing striping artifacts in light-sheet fluorescence microscopy: a review},
  author={Ricci, Pietro and Gavryusev, Vladislav and M{\"u}llenbroich, Caroline and Turrini, Lapo and de Vito, Giuseppe and Silvestri, Ludovico and Sancataldo, Giuseppe and Pavone, Francesco Saverio},
  journal={Progress in biophysics and molecular biology},
  volume={168},
  pages={52--65},
  year={2022},
  doi={10.1016/j.pbiomolbio.2021.07.003},
  publisher={Elsevier}
}

@article{wu2016simultaneous,
  title={Simultaneous multiview capture and fusion improves spatial resolution in wide-field and light-sheet microscopy},
  author={Wu, Yicong and Chandris, Panagiotis and Winter, Peter W and Kim, Edward Y and Jaumouill{\'e}, Valentin and Kumar, Abhishek and Guo, Min and Leung, Jacqueline M and Smith, Corey and Rey-Suarez, Ivan and others},
  journal={Optica},
  volume={3},
  number={8},
  pages={897--910},
  year={2016},
  doi={10.1364/OPTICA.3.000897},
  publisher={Optical Society of America}
}

@article{preibisch2014efficient,
  title={Efficient {B}ayesian-based multiview deconvolution},
  author={Preibisch, Stephan and Amat, Fernando and Stamataki, Evangelia and Sarov, Mihail and Singer, Robert H and Myers, Eugene and Tomancak, Pavel},
  journal={Nature methods},
  volume={11},
  number={6},
  pages={645--648},
  year={2014},
  doi={10.1038/nMeth.2929},
  publisher={Nature Publishing Group US New York}
}

@article{vladimirov2014light,
  title={Light-sheet functional imaging in fictively behaving zebrafish},
  author={Vladimirov, Nikita and Mu, Yu and Kawashima, Takashi and Bennett, Davis V and Yang, Chao-Tsung and Looger, Loren L and Keller, Philipp J and Freeman, Jeremy and Ahrens, Misha B},
  journal={Nature methods},
  volume={11},
  number={9},
  pages={883--884},
  year={2014},
  doi={10.1038/nmeth.3040},
  publisher={Nature Publishing Group US New York}
}

@article{Krzic2012MuViSPIM,
author = {Krzic, Uros and G{\"u}nther, Stefan and Saunders, Timothy E. and
Streichan, Sebastian J. and Hufnagel, Lars},
title = {Multiview light-sheet microscope for rapid in toto imaging},
journal = {Nature Methods},
year = {2012},
volume = {9},
number = {7},
pages = {730--733},
doi = {10.1038/nmeth.2064}
}

@article{Girstmair2022XOpenSPIM,
author = {Girstmair, Johannes and Moon, HongKee and Brillard, Charl{\`e}ne and
Haase, Robert and Tomancak, Pavel},
title = {Time to Upgrade: A New OpenSPIM Guide to Build and Operate Advanced OpenSPIM Configurations},
journal = {Advanced Biology},
year = {2022},
volume = {6},
number = {4},
pages = {2101182},
doi = {10.1002/adbi.202101182}
}

@article{deMedeiros2022MultiscaleOrganoid,
author = {de Medeiros, Gustavo and Ortiz, Raphael and Strnad, Petr and
Boni, Andrea and Moos, Franziska and Repina, Nicole and
Challet Meylan, Ludivine and Maurer, Francisca and Liberali, Prisca},
title = {Multiscale light-sheet organoid imaging framework},
journal = {Nature Communications},
year = {2022},
volume = {13},
pages = {4864},
doi = {10.1038/s41467-022-32465-z}
}

@article{Wysmolek2022MinimalComplexityLSFM,
author = {Wysmolek, Paulina M. and Kiessler, Filippo D. and
Salbaum, Katja A. and Shelton, Elijah R. and
Sonntag, Selina M. and Serwane, Friedhelm},
title = {A minimal-complexity light-sheet microscope maps network activity in {3D} neuronal systems},
journal = {Scientific Reports},
year = {2022},
volume = {12},
pages = {20420},
doi = {10.1038/s41598-022-24350-y}
}

@article{Beghin2022JeWells,
author = {Beghin, Anne and Grenci, Gianluca and Sahni, Geetika and
Guo, Su and Rajendiran, Harini and Delaire, Tom and
Mohamad Raffi, Saburnisha Binte and Blanc, Damien and
de Mets, Richard and Ong, Hui Ting and
Galindo, Xareni and Monet, Anais and Acharya, Vidhyalakshmi and
Racine, Victor and Levet, Florian and Galland, Remi and
Sibarita, Jean-Baptiste and Viasnoff, Virgile},
title = {Automated high-speed {3D} imaging of organoid cultures with multi-scale phenotypic quantification},
journal = {Nature Methods},
year = {2022},
volume = {19},
number = {7},
pages = {881--892},
doi = {10.1038/s41592-022-01508-0}
}

@article{Jain2025BrainOrganoidMorphodynamics,
author = {Jain, Akanksha and Gut, Gilles and Sanchis-Calleja, Fatima and
Tschannen, Reto and He, Zhisong and Luginbuhl, Nicolas and
Zenk, Fides and Chrisnandy, Antonius and Streib, Simon and
Harmel, Christoph and Okamoto, Ryoko and Santel, Malgorzata and
Seimiya, Makiko and Holtackers, Rene and Rohland, Juliane K. and
Jansen, Sophie Martina Johanna and Lutolf, Matthias P. and
Camp, J. Gray and Treutlein, Barbara},
title = {Morphodynamics of human early brain organoid development},
journal = {Nature},
year = {2025},
volume = {644},
number = {8078},
pages = {1010--1019},
doi = {10.1038/s41586-025-09151-3}
}

@article{Moos2024DualViewGastruloids,
author = {Moos, Franziska and Suppinger, Simon and de Medeiros, Gustavo and
Oost, Koen Cornelius and Boni, Andrea and Remy, Camille and
Weevers, Sera Lotte and Tsiairis, Charisios and
Strnad, Petr and Liberali, Prisca},
title = {Open-top multisample dual-view light-sheet microscope for live imaging of large multicellular systems},
journal = {Nature Methods},
year = {2024},
volume = {21},
pages = {798--803},
doi = {10.1038/s41592-024-02213-w}
}

@article{Suppinger2023Gastruloids,
author = {Suppinger, Simon and Zinner, Marietta and Aizarani, Nadim and others},
title = {Multimodal characterization of murine gastruloid development},
journal = {Cell Stem Cell},
year = {2023},
volume = {30},
number = {6},
pages = {867--884.e11},
doi = {10.1016/j.stem.2023.04.015}
}

@dataset{Mossige_2026_zenodo_23898349,
  author    = {Mossige, Endre Joachim Lerheim},
  title     = {From {O}pen{SPIM} to {F}low{SPIM}: electronic material},
  publisher = {Zenodo},
  year      = {2026},
  month     = nov,
  day       = {3},
  doi       = {10.5281/zenodo.23898349},
  url       = {https://doi.org/10.5281/zenodo.23898349},
  note      = {Zenodo record 23898349}
}

\clearpage

\section{Tables and Figures}  
\clearpage

%Figure 1

\clearpage
%Figure 2

\clearpage
%Table 1

\begin{table}[]
    \includegraphics[width=1.0\linewidth]{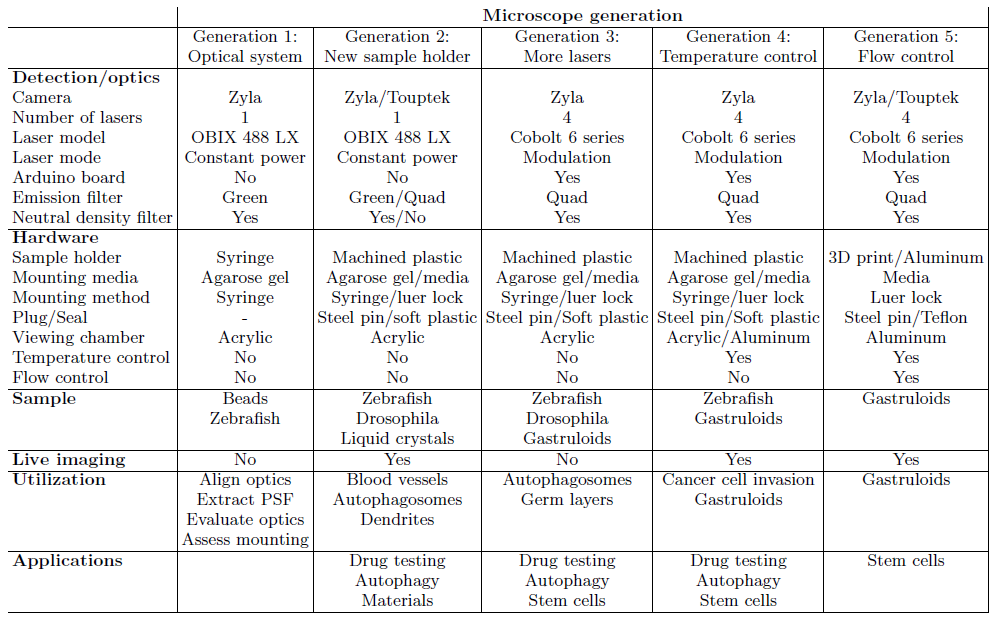}
    \caption{The table gives an overview of the different generations of our setup, from a stripped-down OpenSPIM microscope in Generation 1 to a versatile 3D imaging tool for live tissue cultures in Generation 5. Each generation builds on and learns from the previous one; for each new generation, we improve the hardware and adjust the experimental protocol. Here, the plastic material used in the machine holders in Generations 2 through 4 is polyoxymethylene, the Green LP emission filter is a long-pass filter blocking out the incident wavelength from the OBIS 488 LX laser, and the Quad BP emission filter in Generations 2 through 5 is a bandpass filter with four bands, each optimized to block the four incident wavelengths from the Cobalt 6 series laser. }
    \label{tab:iterativedesign}
\end{table}

\clearpage
%Figure 3

%Figure 4
\clearpage

\clearpage
%Figure 5

\clearpage 
%Figure 6

\clearpage
%Figure 7: Overview of sample holders

\clearpage
%Figure 8: mounting protocol

\begin{figure}[p]
    \centering
    \includegraphics[width=0.4\linewidth]{MOUNTING.jpeg}
    \caption{Mounting protocol. \textbf{Top:} A simple slip-tip syringe and needle system is used to aspirate large specimens, such as zebrafish larvae, into the capillary tube. \textbf{Bottom:} Mounting smaller samples such as gastruloids is more delicate and is performed using a luer-lock system and flanged tubing.}
    \label{fig:mounting}
\end{figure}

%Figure 9: Hardware interfacing
\clearpage

\clearpage
%Figure 10: Multichannel imaging

\clearpage
%Figure 11: Temperature control unit

\clearpage
%Figure 12: FlowSPIM: Overview of flow perfusion system

%\clearpage
%%Figure 13: FlowSPIM: Optical section of a live gastruloid in two channels
%\begin{figure}[h]
 %   \centering
    %\includegraphics[width=0.5\linewidth]{figures/Gen5/Composite_20250613_4dpa_Chi_live_1_v2_scalebar.png}
    %\caption{Optical section of a live gastruloid at 4 days-post-aggregation (4 dpa), imaged 30 minutes after mounting in the capillary tube. The green signal corresponds to T/Bra GFP (mesoderm) reporters, and the red signal corresponds to Sox17 RFP (endoderm) reporters. The light sheet comes from the left. Scale bar: 100\,$\mu$m.}
    %\label{fig:livegastruloid}
%\end{figure}

\clearpage
%Figure 13: FlowSPIM: Optical section of a live gastruloid in two channels (deconvolved in Huygens using a theoretical PSF)

\begin{table}[]
    \includegraphics[width=1.0\linewidth]{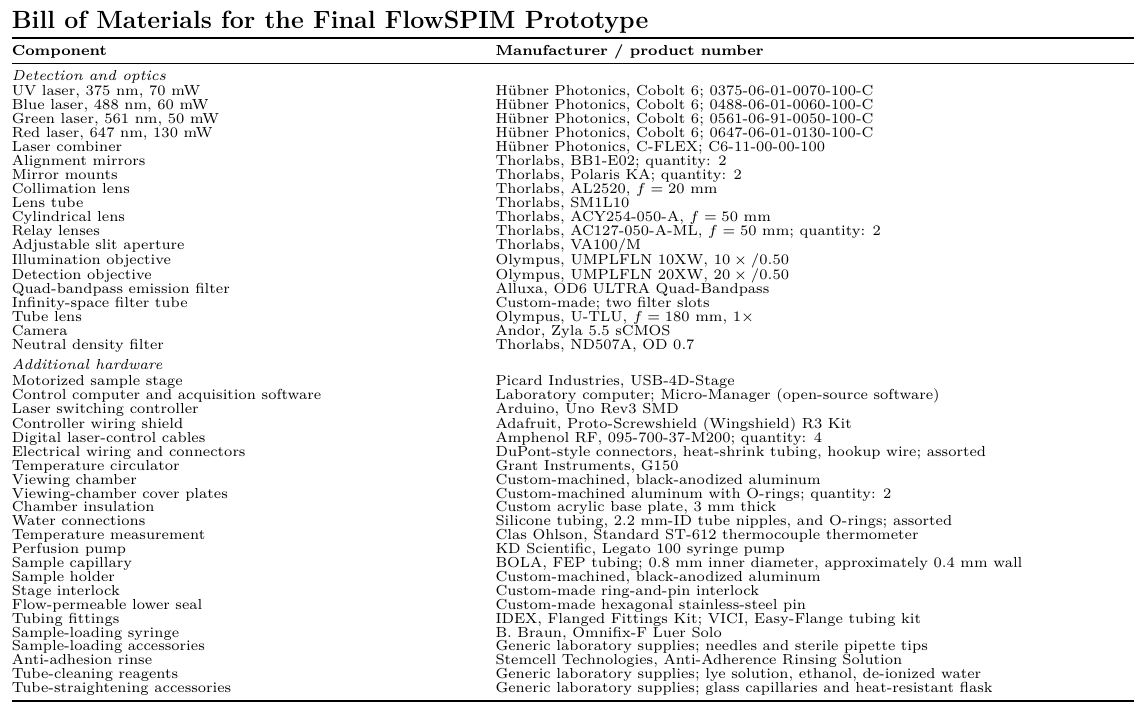}
    \caption{Bill of materials for the FlowSPIM, which is the last prototype of the setup (Prototype 5D).}
    \label{tab:BOM_FlowSPIM}
\end{table}

\end{document}